\documentclass[useAMS,times,mathptm,usenatbib,usegraphicx]{mn2e}

\title[The Fundamental Properties of Galaxies and a New Galaxy Classification System]{The Fundamental Properties of Galaxies and a New Galaxy Classification System}
\author[Christopher J. Conselice]{Christopher J. Conselice$^{1}$\thanks{E-mail:
conselice@nottingham.ac.uk} \\
$^{1}$University of Nottingham, School of Physics \& Astronomy, Nottingham, NG7 2RD UK}

\def\solm{M$_{\odot}\,$}

\def\kms{km s$^{-1}$}

\begin{document} 

\date{Accepted ; Received ; in original form}

\pagerange{\pageref{firstpage}--\pageref{lastpage}} \pubyear{2002}

\maketitle

\label{firstpage}

\begin{abstract}

We present in this paper a new three-dimensional galaxy classification 
system designed to account
for the diversity of galaxy properties in the nearby universe. To construct
this system we statistically analyse a sample of $>22,000$ galaxies at 
$v~<~15,000$ \kms\, ($z < 0.05$) with Spearman rank and principal 
component analyses. Fourteen major galaxy 
properties are considered, including:
Hubble type, size, colour, surface brightness, magnitude, stellar
mass, internal velocities, HI gas content, and an index that measures
dynamical disturbances.  We find, to a high degree, that most galaxy
properties are correlated, with in particular Hubble
type, colour, and stellar mass all strongly related. 
We argue that this tight 3-way correlation is a result of 
evolutionary processes that depend on galaxy mass, as we show that
the relation 
between colour and mass is independent of Hubble type.
Various principal component analyses reveal that most of
the variation in nearby galaxy properties can be accounted for by
eigenvectors dominated by: (i) the scale of a galaxy, such as its
stellar mass, (ii) the spectral type,  and (iii) the degree of dynamical 
disturbances.
We suggest that these three properties: mass, star formation, and
interactions/mergers are the major features that determine a galaxy's physical
state, and should be used to classify galaxies.  As shown in 
Conselice (2003), 
these properties are measurable within the CAS (concentration, asymmetry, 
clumpiness) structural system, thus providing an efficient mechanism for 
classifying galaxies in optical light within a physical meaningful framework.  
We furthermore discuss the fraction and number
density of galaxies in the nearby universe as a function of Hubble
type, for comparison 
with higher redshift populations.

\end{abstract}


\section{Introduction}


Galaxies as distinct stellar, gaseous, and dark matter units have dozens of
properties measurable by a variety of techniques from gamma-ray to radio
wavelengths.  
Some of the most basic properties include: spectral energy
distributions, spectral line properties, sizes, internal velocities, 
structural features, and morphologies.
From these observable quantities we can derive additional
physical information, such as stellar, gaseous, and total masses, 
metallicities, the current and past star formation rates, and the fraction 
of light in various structural components.  Each of these features reveals 
potentially important clues for how galaxies, including their gas content
and dark matter
halos, were created and have evolved.  These properties are also interrelated
through well-known scalings, such as the
colour-magnitude relation (e.g., Sandage \& Visvanathan 1978), the 
size-magnitude relation (e.g., Romanishin 1986), the
Tully-Fisher relation (Tully \& Fisher 1977),
and the Fundamental Plane (Djorgovski \& Davis 1987; Dressler et al. 1987). 
These correlations all describe the scaling of 
structural features of normal galaxies with each other, or with the 
internal kinematics of galaxies. 

It is however not yet clear what the most basic 
evolutionary processes and galaxy observables
are, and whether or not well known relations 
between galaxy parameters capture essential galaxy features.
Is there a single, or a group of processes responsible for
the diversity of galaxies?  Although it is widely known that
physical properties vary along the Hubble sequence,  
galaxies within a Hubble type have a large range of sizes, colours, 
surface brightnesses, masses, and so on. Perhaps there is an alternative
way to classify galaxies based on the properties responsible for
driving evolution.   One goal of
this paper is determining what physical features of galaxies are the
best for classification purposes.  Currently it is assumed by many that 
the optical 
morphology of a galaxy provides this classification, either in the form
of a Hubble type, or through quantitative methods.    However there
is no a priori reason that optical morphologies should reveal information
needed to classify galaxies into a meaningful system. 

We utilise fourteen properties of 22,121 galaxies in the nearby
universe to address this problem. We begin our analysis with 
the hypothesis that our fourteen measurable 
features provide the basic physical information needed to fully
characterise nearby galaxies.
We analyse these measurables through statistical methods, such as 
Spearman rank and principal component analyses, to determine which observables
are the most
fundamental for classifying galaxies. We demonstrate that at a fundamental
level many measured properties of galaxies correlate with each other.   
For example, scale features correlate strongly, and Hubble 
types correlate with many properties, including colour and stellar mass.  We
argue that this correlation is the reason that Hubble types have
been successful in describing the nearby galaxy population.

We further show through principal component analyses that most of the 
variation in galaxy properties can be accounted for by three eigenvectors 
dominated by scale (e.g., mass), the presence 
of star formation, and the degree of recent interaction/merger activity.
As a result, our main conclusion is that these 
three features are needed  to understand 
the current and past evolutionary state of a galaxy. 
We further argue that galaxy mass, star formation, and mergers/interactions 
are responsible for the diversity in the galaxy population in
the nearby universe, and should be used to construct a galaxy
classification system.

This paper is a more general companion, and in some ways a prequel,
 to the quantitative 
CAS (concentration, asymmetry, clumpiness) structural analysis paper 
presented in Conselice (2003).  In Conselice (2003) we created a new
quantitative way to classify galaxies under the assumption that mass, 
star formation, and recent
mergers were the critical properties for understanding galaxies. We now show
that this assumption is justifiable through our statistical analysis
of $> 22,000$ nearby galaxies.  Conselice
(2003) furthermore developed a quantitative and reproducible  method for 
retrieving this
information through the optical light of galaxies.

The outline of this paper is as follows.  In \S 2 we give a summary of
the physical features, and formation histories of various types of galaxies
we study.  \S 3 describes
our galaxy sample, and the properties we use to characterise each system. \S 4
gives a detailed description of the nearby galaxy population, including
the relative numbers and number densities of various morphological types.
\S 5 presents our statistical analysis of our sample, while \S 6 utilises this
information to produce a new physical classification system, and \S 7
is a summary.  For distance measurements we use a Hubble constant of
H$_{0} = 70$ km s$^{-1}$ Mpc$^{-1}$, unless otherwise noted.

\section{Nearby Galaxy Types and their Possible Formation Histories}

There are a few generally agreed upon major 
nearby basic galaxy types, generally classified according to 
morphology, whose properties and range of
possible formation histories are becoming well understood. 
These types are: ellipticals, spirals, lenticulars, dwarf 
spheroidals/ellipticals, low surface-brightness galaxies, irregulars, and 
mergers, which we review below.   These groups can, but do not
necessarily include, other types of galaxies such as starbursting
systems (e.g., ultra luminous infrared galaxies, ULIRGs), galaxies
with active galactic nuclei (quasars, radio galaxies, etc.).

Most {\bf ellipticals}, particularly those in clusters, contain old stellar 
populations that likely formed early in the universe (Kuntschner \& Davies
1998).   Many of these galaxies are thought to form by the 
mergers of smaller galaxies (Barnes \& Hernquist 1992), 
or from a rapid collapse of non-rotating gas (van
Albada 1982).  The merger idea for the origin of ellipticals
is currently popular, with both
indirect and direct supporting evidence (e.g., Schweizer \& Seiter 1988; 
Conselice 2006).

{\bf Spiral galaxies} are a mixed galaxy class, but all generally show
evidence of recent star formation and past merger/accretion 
activity (e.g., Kennicutt 1998; Zaritsky \& Rix 1997). The stellar populations
in spirals are also younger, on average, than those found in giant
ellipticals (e.g, Trager et al. 2000).

The {\bf lenticular class (S0s)} are disk-like galaxies without the presence 
of strong
spirals arms, star-formation, obvious dust, or gas usually associated
with spirals and other star-forming galaxies.   S0 galaxies
are usually found in galaxy clusters (Dressler 1984), but can also
exist in the field (e.g,. van den Bergh 1997).
 It has been suggested since Spitzer \& Baade (1951)
that S0s originate from spirals due to environmental effects.
Evidence for this includes spirals in nearby
clusters that are highly depleted in gas (Haynes \& Giovanelli 1986), and 
might be evolving into S0 galaxies.  

{\bf Dwarf ellipticals (dE)}, the most common type of galaxy in the nearby
universe, are almost always located in dense environments such as galaxy 
clusters and groups (Ferguson \& Binggeli 1994).   These 
objects resemble ellipticals due to their
symmetric structure, and elliptical isophotes, although they differ in 
several significant
properties, such as containing different light profiles, and 
how they follow scaling relations (e.g., Wirth \& Gallagher 1984; Graham
\& Guzman 2003). These
objects likely have several formation mechanisms, including
a primordial origin and evolution from stripped galaxies (Conselice
et al. 2001,2003c).

{\bf Low surface brightness galaxies (LSBs)} are one of the last major class
 of galaxies discovered to date (Disney 1976).  These objects come in many 
shapes
and sizes, but generally mimic the Hubble sequence in terms of
bulge to disk ratios, but at a much lower
surface brightness, at $\mu$ $>24$ mag arcsec$^{-2}$ 
(see Impey \& Bothun 1997). 
It has been suggested from several pieces of observational
evidence that LSB galaxies have had few, if any, major merging or
interaction events with other galaxies, and are found in relative isolation
(e.g., Knezek 1993; Mo, McGaugh \& 
Bothun 1994).  LSB galaxies therefore potentially represent a class of 
objects formed mostly through the collapse of gas, with gradual and continual 
ongoing star formation, with few interactions throughout their history.   

{\bf Irregulars} are typically smallish star forming galaxies that often 
lack a distinct 
spiral structure.  For our purposes, irregulars include dwarf irregulars,
blue compact dwarfs, and classical irregulars such as NGC 4449.
Star formation dominates the appearance and evolution of
these small, but common, galaxies that have a wide
range of surface brightnesses
(see Gallagher \& Hunter 1984 for a review).  They are also the bluest
type of non-interacting galaxy known (Huchra 1977).
A large amount of attention has been spent studying the dIrr members of
the local group, although these objects also exist in clusters (e.g., 
Binggeli, Sandage \& Tammann 1985; Gallagher \& Hunter 1986).

Finally, {\bf interacting and merging galaxies}, while rare in the 
nearby universe, were much more common in the distant past
when roughly half of all massive galaxies were undergoing major mergers (e.g.,
Conselice et al. 2003).    Mergers are likely a critically important
galaxy formation/evolution process especially if the first galaxies
were all of low mass. Mergers are also observationally more common at earlier
times (e.g., Conselice et al. 2003a, Bundy et al. 2004, Lin
et al. 2004, Pope et al. 2005).

\section{Nearby Galaxy Analysis}

\subsection{RC3 Sample and Data}

The sample of galaxies with which we performed our statistical analyses 
were taken from galaxies in the Third Reference Catalog of
Bright Galaxies (hereafter RC3; de Vaucouleurs et al. 1991).  These 
RC3 galaxies were chosen based on their size ($> 1$\arcmin),
apparent magnitude B$_{\rm T} < 15.5$ and radial velocity (v $< 15,000$
\kms), although the catalogue is supplemented by objects that met only
one of these criteria.  This catalog is however inhomogeneous, which we
deal with in several ways, described in the following analysis.
For each galaxy  we culled 
from the literature, or derive, the following features.  Our total
sample consists of 22,121 galaxies distributed across the sky as
shown in Figure~1.  The average galaxy in our sample is at L$_{*}$, and
its average galaxy colour is $(B-V)_{0} = 0.66$.

\begin{figure}
\includegraphics[width=84mm]{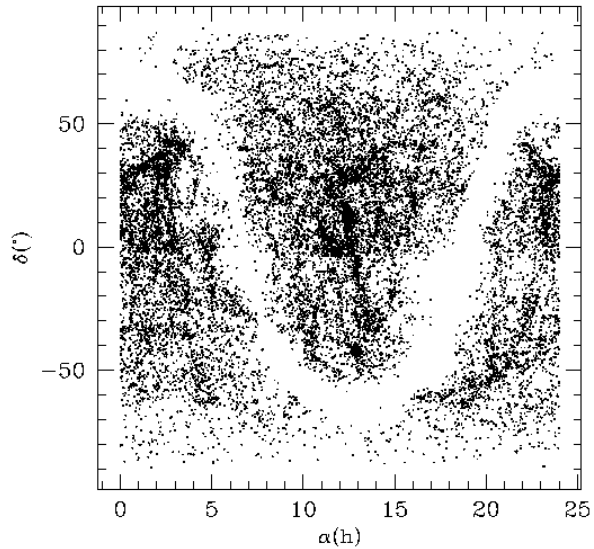}
 \caption{The spatial distribution of the galaxies used for our statistical
study of galaxy populations and their properties.  The U-shaped gap is
the galactic plane, and the structure at the centre constitutes the
Virgo supercluster. }
 \label{sample-figure}
\end{figure}

\noindent {\bf I. T-Type (T)}:  The mean de Vaucouleurs (1959) T-type is an 
expression of the gross morphology of a galaxy.  The T-type values we used are 
means for each galaxy, computed from a variety of sources (RC3).  These T-types
can be equated with Hubble types as: -6 through -4 are ellipticals, 
-3 through -1 are S0s, and (0,1,2,3,4,5,6,7,8,9) are (S0/a, Sa, Sab, Sb, 
Sbc, Sc, Scd, Sd, Sdm, Sm), respectively.  The RC3 is the largest eye-ball
estimated morphological catalogue to date.

\noindent {\bf II. Size (R)}: The sizes of each galaxy is computed
from the isophotal major axis diameter (D$_{25}$) measured in arc-minutes and
converted into kpc by using the distance modulus for each galaxy.  While
Petrosian radii, or a radius which does not use isophotes, is more ideal,
the effects of using an isophotal radii for these nearby
systems is minor.

\noindent {\bf III. Mean Effective Surface Brightness ($<\mu>_{e}$)}: The mean
effective surface brightness is the surface brightness within the half light
radius of each galaxy.  This parameter, effectively combines the
size and magnitude of a galaxy and gives us a rough idea of the projected
density of stars.

\noindent {\bf IV. $(U-B)_{0}$ and $(B-V)_{0}$ colours}: 
these are photometrically
determined colour indices corrected for galactic and internal extinction.
We used these indices as a measure of the average luminosity weighted age 
of 
stars in galaxies, although for older ($> 6$ Gyr) stellar populations 
these colours can
be dominated by metallicity effects (e.g., Worthey 1994).

\noindent {\bf V. HI linewidths (W$_{20}$ and W$_{50}$)}: The HI line
widths at the 20\% and 50\% of the maximum level in the 21-cm line profile.

\noindent {\bf VI. HI linewidth ratio ($I$ = W$_{20}$/W$_{50}$)}.  We used 
this number, based on the arguments in
Conselice, Bershady \& Gallagher (2000b) as a measure of the importance of
recent interactions and mergers.  Values of I$>1.5$ are
interpreted as systems that have recently undergone interactions
or merger with other galaxies.  We discuss this index in more detail
in \S 3.2 and Conselice et al. (2000b).

\noindent {\bf VII. Central velocity dispersion (V$_{\rm S}$)}.  
This is usually only
defined for early type galaxies, and V$_{\rm S}^{2}$
times a size can be used as a measure of early type galaxy mass.

\noindent {\bf VIII. HI line index (HI)} is defined
as m$_{21}$ = 16.6 - 2.5 log $S_{H}$, where $S_{H}$ is the flux of the
HI line measured in units of
W m$^{-2}$.   We further subtract this value by the distance modulus (DM)
to obtain physical quantities that are independent of distance.  We use the
HI line index as a measure of the cold neutral gas content in a galaxy.

\noindent {\bf IX. Far infrared index (FIR)} is defined in a similar manner
as the HI line index (HI), as m$_{\rm FIR} = -20 - 2.5$ log(FIR) - DM,
where FIR = 1.26[2.58 $\times f_{\nu}(60 \mu {\rm m}) + f_{\nu}(100 \mu
{\rm m})$], and f$_{\nu}(60 \mu {\rm m})$ and f$_{\nu}(100 \mu
{\rm m})$ are the total flux densities in the IRAS 60 and 100 $\mu$m
bands (RC3) and where DM is the distance modulus of the galaxy.

\noindent {\bf X. Luminosity class ($\lambda$)}: $\lambda$
is the van den Bergh (1966) DDO, morphologically determined luminosity
class index, which have been extended to fainter magnitudes
in the RC3.  The luminosity class values we used are weighted averages, just
as the T-types, and have values that range from $\lambda = 1$ to 11 with
higher $\lambda$ values for fainter galaxies.

\noindent {\bf XI. The maximum rotational velocity (V$_{\rm max}$)}:  
V$_{\rm max}$ is
defined as the maximum rotational velocity of a galaxy, as measured
from HI or H$\alpha$ rotation curve profiles.  The size of a disk galaxy
times V$_{\rm max}^{2}$ is sometimes used as a measure of its total mass.

\noindent {\bf XII. The absolute B-band magnitude (M$_{\rm B}$)}. 

\noindent {\bf XIII.  The environmental galaxy density $\rho$}.  This is 
defined as the number of
galaxies within 3 Mpc and $< 1000$ \kms\, of each galaxy in our catalog.
To avoid severe incompleteness for faint galaxies, we only compute this
number for bright galaxies with M$_{\rm B} < -20$ and at declinations
$\delta > -30$.

\noindent {\bf XIV.  Galaxy Stellar Mass}.  Stellar masses for our
sample were computed by deriving the stellar (M/L) ratio in the B-band
using the (B$-$V)$-$M/L ratio relation discussed in Bell et al. 
(2003), using their `diet' Salpeter initial mass function.  
There is a general correlation 
between stellar (M/L) and colour, such that bluer colors have lower stellar
(M/L) ratios.   
We used this relation, and the measured $(B-V)_{0}$ 
colours for our galaxies to derive
the (M/L) in the B-band for our sample. We then compute stellar masses by
multiplying the luminosity in the B-band for each galaxy by its derived
stellar (M/L) ratio.

Not all galaxies in our sample  have
each of these parameters measured, therefore when comparing different
parameters we often had to use a subset of the data.  We also associate
with these parameters an error that reflects the variation in the
measured values.

\subsection{The Galaxy Interaction/Merger Index ($I$)}

The galaxy interaction and merger index, $I$, defined by the
ratio of the width of the HI line at 20\% and 50\% of maximum 
(property VI in \S 3.1) is used in this paper as the principle indicator
for finding galaxies undergoing mergers and interactions.  It is
a dynamical indicator and has been used at least once before
for this purpose in Conselice et al. (2000b).  However, it is not
a standard galaxy measure and it is worth describing it, and the
galaxies it selects, in more detail.

The idea behind its use is that a normal galaxy with HI, will present
a regular HI profile that is relatively narrow, or at least does not
significantly change shape.   If a galaxy is interacting or merging, and HI gas
is disturbed by these dynamical events then the HI profile will
have a shallow rise, or will have wings.  This leads to higher values of
$I$ =W$_{20}$/W$_{50}$.  Conselice et al. (2000b) show that the $I$ 
index correlates well with the asymmetry parameter, such that galaxies 
with very large asymmetries due to merging have large $I$ values.

\begin{figure}
\includegraphics[width=84mm]{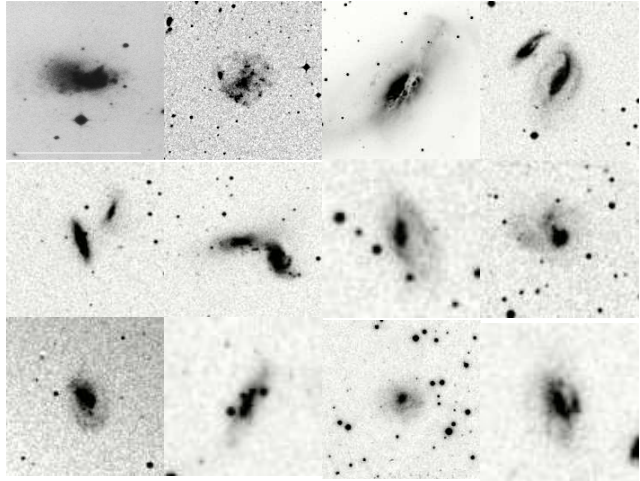}
 \caption{Examples of galaxies within our sample, mostly taken from the 
DSS, which have
large 20\% to 50\% HI line width ratios, $I$.  As can be seen
galaxies in interactions, in pre-mergers and post-mergers will
all produce high HI line width ratios. }
 \label{sample-figure}
\end{figure}

However, beyond the sample in Conselice et al. (2000b), this index
has not been tested to determine its success rate in finding merging
galaxies.  To understand this better we examined the literature and
Digitised Sky Survey (DSS) imaging for several hundred systems 
with $I > 2$.  What we found was a diversity of galaxy morphology, 
from spirals and obvious mergers, to galaxies classified as 
S0s/ellipticals. However, it was nearly always the case that a 
system with a regular apparent morphology was in a galaxy pair, 
such that it is likely
involved in an interaction with another galaxy.  Examples of
the optical morphology of the high-$I$ systems taken from the NASA/IPAC 
Extragalactic Database (NED) and the DSS are shown in Figure~2. It is
clear from these images that the $I$ indicator is a more sensitive
probe than the asymmetry index (Conselice 2003), which is designed to
find galaxies that have already merged. It appears that the $I$ index
will become large during an interaction as well as throughout 
the merger of two galaxies.  There is however some contamination from
non-interacting galaxies into our high $I$ sample, as might be expected.  
This rate is however low, and does not affect our results, as it is
only at the 5\% level. Much of this contamination seems to arise from edge-on 
disk galaxies.

\section{Nearby Galaxy Properties}

Our primary goal using our large sized-limited sample of nearby galaxies with 
morphologies is determining the strongest 
correlations between measurable parameters (\S 5).  Before doing
this, we examine the  nearby galaxy 
population, its overall distribution of properties, and how each
relates to morphological type.   One purpose for this is having a well 
defined morphological census of nearby galaxies to compare with more
distant samples studied with the Hubble Space Telescope.  

\begin{figure}
\includegraphics[width=84mm]{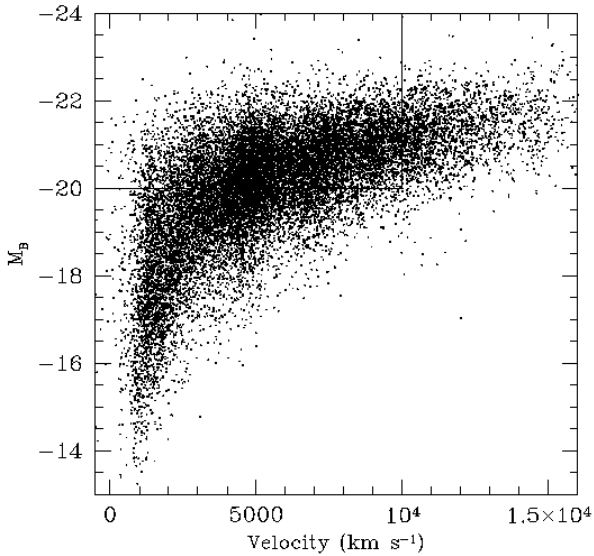}
 \caption{The radial velocity distribution of the galaxies used in our
analysis verses their absolute magnitude (M$_{\rm B}$). The box formed
by the solid line outlines the nearly complete sample we used for analyses
that require near complete samples brighter than M$_{\rm B} < -20$.}
 \label{sample-figure}
\end{figure}

\subsection{Galaxy Stellar Mass and Spectro-Morphology at $z \sim 0$}

As the Hubble Space Telescope can now reveal the morphological properties, 
and stellar masses of
thousands of galaxies at $z > 0.5$ (Papovich et al. 2005; Bundy
et al. 2005; Conselice et al. 2004, 2005a), direct morphological and
stellar mass comparisons with the $z = 0$ galaxy population are now
possible (e.g., van den Bergh, Cohen \& Crabbe 2001; Conselice et al. 2005). 
It is thus desirable to determine the morphological
distribution of galaxies at $z \sim 0$ and how these
relate to stellar masses and colours.

\setcounter{table}{0}
\begin{table*}
 \caption{Morphological Fractions and Densities of Nearby Bright Galaxies}
 \label{tab1}
 \begin{tabular}{@{}cccccccccccc}
  \hline
M$_{\rm B} = -22$ & (\# Gpc$^{-3}$ h$_{100}^{3}$) & Fraction &  & M$_{\rm B} = -21$ & (\# Gpc$^{-3}$ h$_{100}^{3}$) & Fraction &  & M$_{\rm B} = -20$ & (\# Gpc$^{-3}$ h$_{100}^{3}$) & Fraction \\
\hline
E & 2.0$\pm0.8$ $\times 10^{4}$ & 0.19 & & E     & 1.3$\pm0.4$ $\times 10^{5}$ & 0.09 & & E & 2.7$\pm0.8$ $\times 10^{5}$ & 0.07 \\
S0 & 1.1$\pm0.5$ $\times 10^{4}$ & 0.10 & & S0 & 1.6$\pm0.5$ $\times 10^{5}$ & 0.12 & & S0 & 5.7$\pm1.5$ $\times 10^{5}$ & 0.15 \\
eDisk & 2.5$\pm1.0$ $\times 10^{4}$ & 0.25 & & eDisk & 4.8$\pm1.3$ $\times 10^{5}$ & 0.36 & & eDisk & 1.3$\pm0.3$ $\times 10^{6}$ & 0.36 \\
lDisk & 4.5$\pm1.6$ $\times 10^{4}$ & 0.44 & & lDisk & 5.4$\pm1.5$ $\times 10^{5}$ & 0.41 & & lDisk & 1.5$\pm0.4$ $\times 10^{6}$ & 0.39 \\
Irr & 2.4$\pm1.8$ $\times 10^{3}$ & 0.02 & & Irr   & 2.0$\pm0.8$ $\times 10^{4}$ & 0.02 & & Irr & 8.4$\pm2.7$ $\times 10^{4}$ & 0.02 \\
\hline
 \end{tabular}
\end{table*}

We first calculate some basic spectro-morphological properties of nearby
galaxies,
including how stellar masses relate to morphology at $z \sim 0$.
Our catalog of 22,121 RC3 galaxies
is however inhomogeneous, particularly for low surface brightness systems. 
As Figure~3 shows, we sample fainter galaxies at lower radial velocities, 
implying we are incomplete at fainter magnitudes.  To partially overcome the 
incompleteness of our catalog, 
we limit our morphological statistical analyses to systems which have 
M$_{\rm B} < -20$ and
$v < 10,000$ \kms ($z \sim 0.033$, 133 Mpc) (shown by the lines in
Figure~3).  When investigating fainter 
galaxies with M$_{\rm B} < -18$, we limits our analyses to objects with 
velocities $v < 3000$ \kms ($z \sim 0.01$, 40 Mpc).

\begin{figure*}
 \vbox to 120mm{
\includegraphics[angle=0, width=104mm]{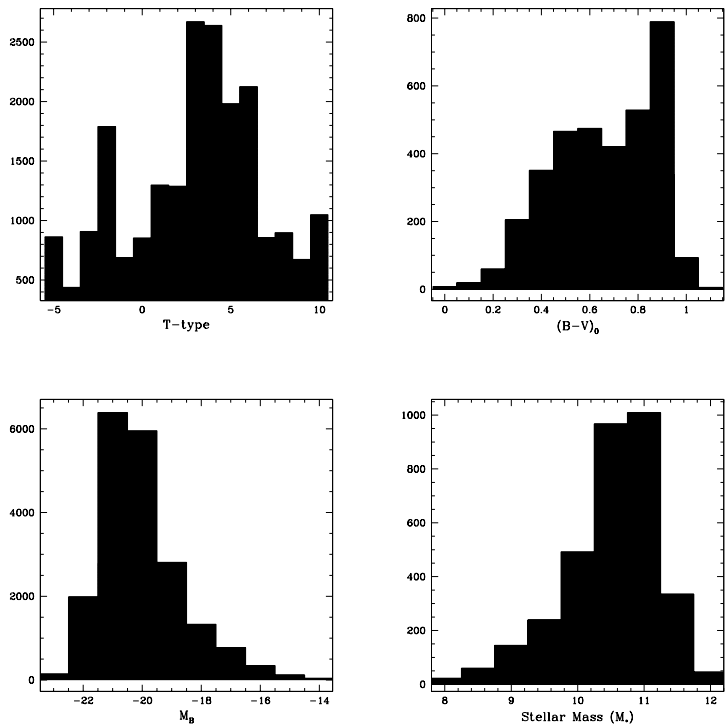}
 \caption{The distribution of morphology (T-type), colour, absolute magnitude,
and stellar mass (M$_{*}$) for
all galaxies used in this study.    We rapidly 
become incomplete at M$_{\rm B} > -20$ and at $M < 5 \times 10^{10}$ \solm
within our full volume (see Figure~3).  As can be seen, we span the entire
range of Hubble types and galaxy colours. }
} \label{sample-figure}
\end{figure*}

In addition to the M$_{\rm B} < -20$ and $v < 10,000$~\kms\, cut, 
we furthermore only consider galaxies 
between declination -50 $< \delta <$ 50, and take into account
the incompleteness of the galaxy distribution due to the Milky Way
(Figure~1) when calculating total number densities. Note that
these number densities are based on the RC3 selection, and
are not complete in any sense.  Our densities however generally 
agree well with the densities derived from the morphological-type specific
Schechter functions published in Marzke et al. (1998) 
and Nakamura et al. (2003).  We investigate
the completeness of our sample down to M$_{B} = -20$ through statistical
tests.  From this sample we create V/V$_{\rm max}$ diagrams
for all morphological types, revealing that we are nearly complete to 10,000
\kms\, for galaxies brighter than M$_{\rm B} = -20$. 

We show fractional histograms in Figure~4 of the distribution of
morphological types, $(B-V)_{0}$ colours, stellar masses, and absolute 
magnitudes (M$_{\rm B}$) for the entire sample.   Figure~5 shows the 
fraction of all galaxies that are classified 
into different T-types. 
From Figure~5, the most common galaxy type in the nearby universe 
appears to be mid/late type spirals, with a peak fraction at
T-type = 4 (Sbc).  This peak remains unchanged when we cut the 
sample at brighter, M$_{B} < -22$, or fainter, M$_{B} < -18$, limits. 

The relative percentages of
nearby galaxies with different morphological types are listed in Table~1.
It appears that $\sim 75$ \% of all nearby
bright galaxies with M$_{\rm B} < -20$ are spirals, with $\sim 40$\%
late type spirals (Sb-Sc).   Thus, in the nearby 
universe most bright galaxies (with M$_{\rm B} < -20$) are spirals with 
small bulges. 
After accounting for spatial incompleteness in the catalog, we further find 
that the total number density of spirals is $\sim 3 \times 10^{6}$
Gpc$^{-3}$ at M$_{\rm B} < -20$ (Table~1).  

\begin{figure}
\includegraphics[width=84mm]{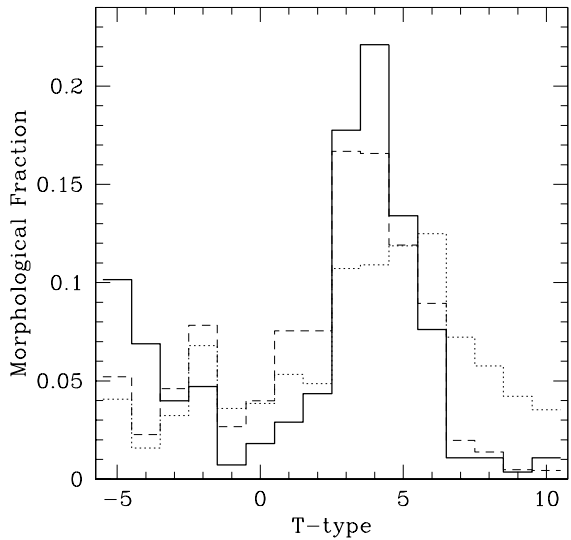}
 \caption{The fraction of different morphological types for all nearby bright 
galaxies with M$_{\rm B} < -22$ (solid line), M$_{\rm B} < -20$ (dashed
line) and M$_{\rm B} < -18$ with $V < 3000$ \kms (dotted line).  The most
common galaxy in our sample is therefore a mid-type spiral, which does not
change significantly at different luminosity cuts.}
 \label{sample-figure}
\end{figure}

Early-type galaxies, including ellipticals and
S0s, account for nearly all of the remainder (22\%) of nearby bright 
galaxies,
with the rest $\sim$ 2\% irregulars (Table~1).  These number densities
and morphological type fractions are being directly compared to bright 
galaxies at high redshifts where these same properties can now 
be determined (e.g., Conselice 2004; Conselice et al. 2005a). 
Figure~6 shows
the fraction of each morphological classed type that is also classified
as having a bar, ring or part of a multiple system, which we discuss below.

\setcounter{table}{1}
\begin{table*}
 \caption{Bar, Ring and Multiple Component Fractions as a Function of
Morphology for Galaxies with M$_{\rm B} < -20$}
 \label{tab1}
 \begin{tabular}{@{}cccccccccccc}
& \underline{Bar}  &  &  & & \underline{Ring} & & & & \underline{Multiple} &  & \\
Type & (\# Gpc$^{-3}$ h$_{100}^{3}$ ) & Fraction &  & Type & (\# Gpc$^{-3}$ h$_{100}^{3}$) & Fraction &  & Type & (\# Gpc$^{-3}$ h$_{100}^{3}$) & Fraction & \\
\hline
E & 1.2$\pm0.5$ $\times 10^{4}$ & 0.01 & & E     & 9.6$\pm10$ $\times 10^{2}$ & 0.00 & & E & 3.0$\pm1.1$ $\times 10^{4}$ & 0.11 \\
S0 & 1.2$\pm0.4$ $\times 10^{5}$ & 0.10 & & S0 & 3.6$\pm1.4$ $\times 10^{4}$ & 0.09 & & S0 & 3.3$\pm1.3$ $\times 10^{4}$ & 0.13 \\
eDisk & 5.3$\pm1.4$ $\times 10^{5}$ & 0.44 & & eDisk & 2.0$\pm0.6$ $\times 10^{5}$ & 0.52 & & eDisk & 9.1$\pm2.9$ $\times 10^{4}$ & 0.35 \\
lDisk & 5.2$\pm1.4$ $\times 10^{5}$ & 0.43 & & lDisk & 1.4$\pm0.4$ $\times 10^{5}$ & 0.37 & & lDisk & 9.3$\pm3.0$ $\times 10^{4}$ & 0.36 \\
Irr & 2.3$\pm0.9$ $\times 10^{4}$ & 0.02 & & Irr   & 4.7$\pm3.0$ $\times 10^{3}$ & 0.01 & & Irr & 1.2$\pm0.6$ $\times 10^{4}$ & 0.05 \\
\hline
 \end{tabular}
\end{table*}

Another feature we investigate is how galaxy morphology correlates
with the positions of galaxies in the colour-magnitude plane (Figure~7).  
There is a well characterised red-sequence and blue-cloud in the 
$(U-B)_{0}$ vs. M$_{\rm B}$ diagram (e.g., Baldry et al. 2004), which has
been proposed as a fundamental feature for understanding how galaxies have 
evolved.   When we label the 
morphological types of galaxies on the colour-magnitude digram we find a
good correlation between the morphologies
of galaxies and their positions.  The early
types (denoted by solid red squares) are mostly on the red-sequence, while
later types (T-type $> 3$, blue crosses) are in the blue cloud, and mid-types
(T-type $> 0$ \& T-type $< 3$) are in between.

\begin{figure*}
 \vbox to 120mm{
\includegraphics[angle=0, width=180mm]{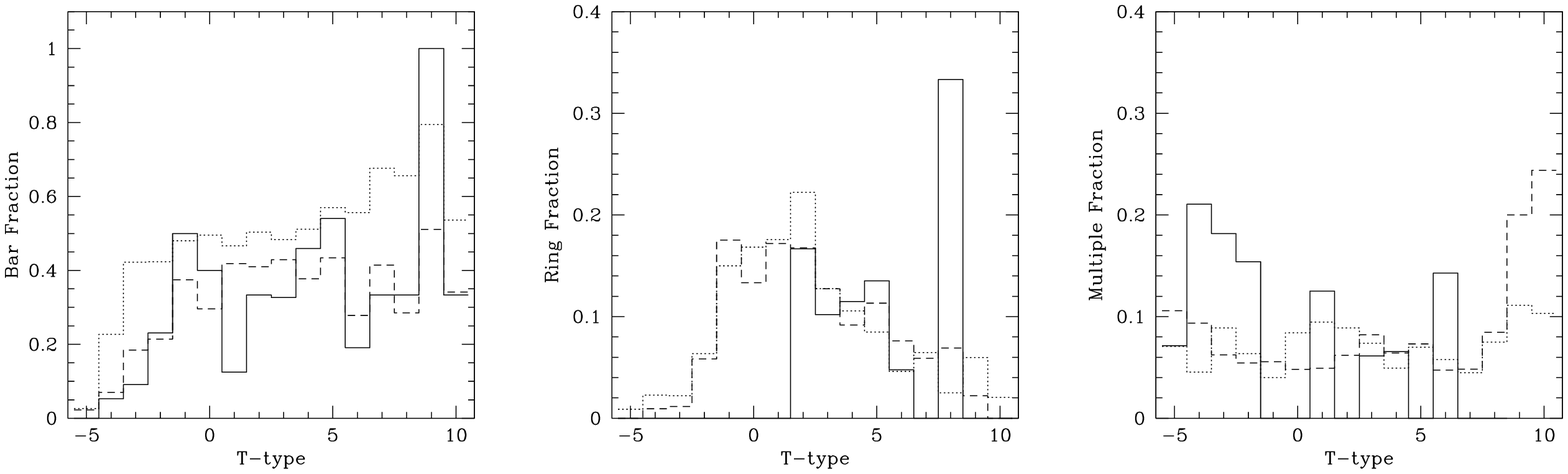}
 \caption{The fraction of different galaxy types that have morphological
evidence for the presence of bars, rings, and multiple components.  These
fractions are also plotted as a function of luminosity, with M$_{\rm B} < 
-22$ (solid line), M$_{\rm B} < -20$ (long-dashed line) and M$_{\rm B} < 
-18$ with $V < 3000$ \kms (dotted line). As can be seen, the bar
fraction is roughly constant at T-types $> 0$, while ringed galaxies are
typically early type spirals, and galaxies with a multiple component
classification tend to span all Hubble types.}
} \label{sample-figure}
\vspace{-3.5cm}
\end{figure*}

This is
also the case for stellar masses, with the M$_{*} > 10^{11}$ \solm
galaxies mostly ellipticals on the red-sequence. Lower mass galaxies are
increasingly bluer. This can also be seen in Figure~8, where we plot the
morphological fraction as a function of stellar mass.  At the highest
masses, nearly all galaxies are ellipticals, but this gradually
changes, and by M$_{*} = 10^{11}$ \solm ellipticals and spirals
each represent about $\sim$50\% of the population.  At the low
mass end, irregulars begin to dominate the population, with
spirals accounting for only 20\% of the population at M$_{*} < 10^{9}$ \solm
and ellipticals accounting for about 10\%. 

\begin{figure}
\includegraphics[width=84mm]{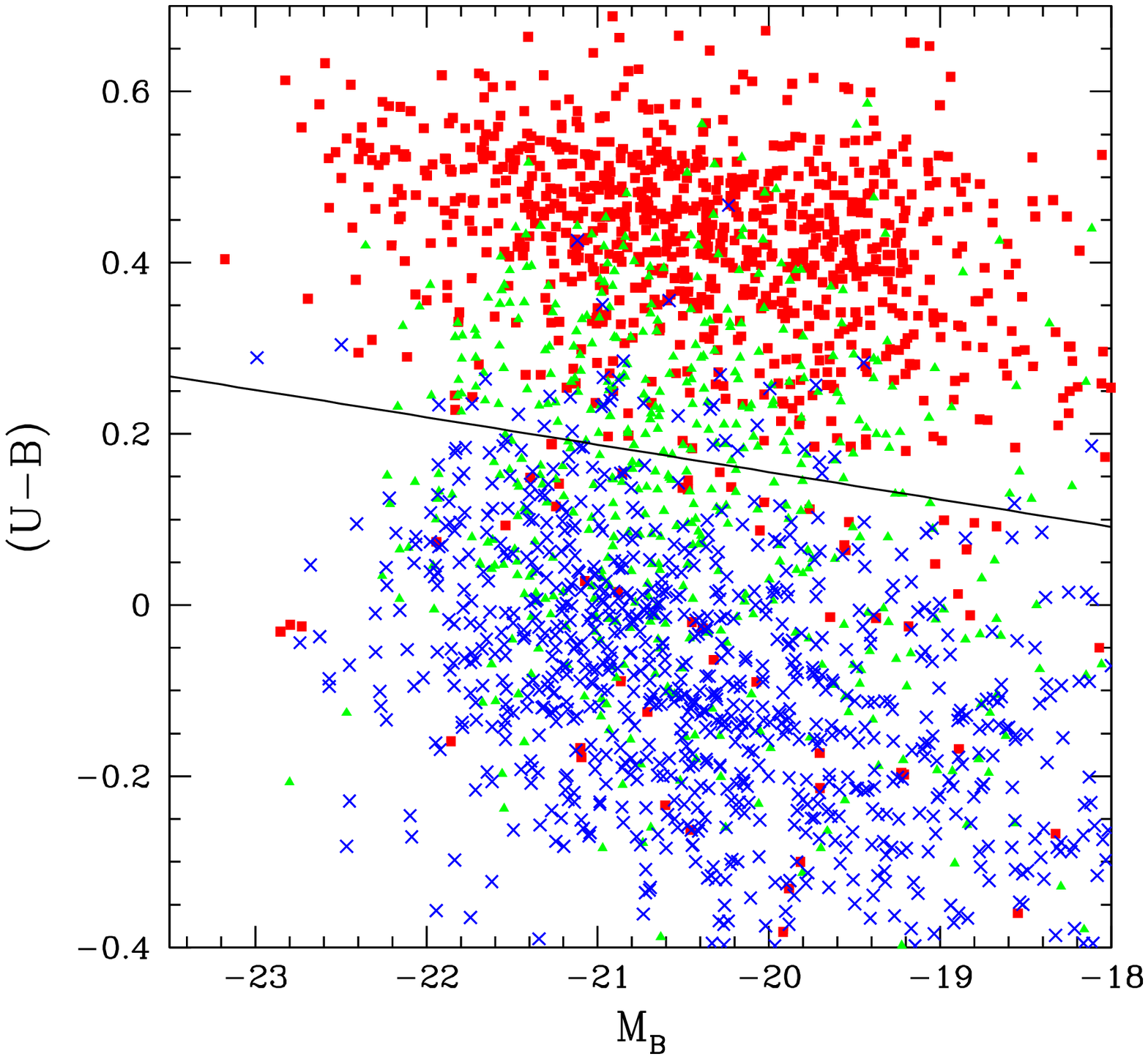}
 \caption{The colour-magnitude relation for galaxies in our sample.
The solid line divides red-sequence galaxies from blue cloud galaxies (Faber
et al. 2005). Early types are denoted by solid red squares, while
later types (T-type $> 3$) are labelled as blue crosses, and mid-types
(T-type $> 0$ \& T-type $< 3$) are labelled as green triangles.}
 \label{sample-figure}
\end{figure}

\subsubsection{Bars}

Bars and rings are usually, but not uniquely, associated with spiral 
galaxies (e.g., Buta \& Combes 1996).  Bars are dynamical
features that reveal, and contribute to, several galaxy evolution processes
such as star formation and AGN activity 
(Barnes \& Hernquist 1992), or galaxy interactions.
It is debated however
whether bars are a fundamental feature of galaxies that should
be used in classifications (e.g., Abraham \& Merrifield 2000; 
Conselice 2003).  The evolution of bars
through cosmic time is also debated, although there are some claims that the
bar fraction and sizes of disks have not evolved since $z \sim 1$ 
(Jogee et al. 2004; Ravindranath et al. 2004).  
As the evolution of bars will continue to be 
addressed in the future through observations with the Hubble Space 
Telescope, we compute basic physical information on barred galaxies in the 
nearby universe for comparison purposes (Figure~6a, Table~2).

Table~2 lists, for all barred galaxies in the nearby universe, the fractional
breakdown of host galaxy morphological type, and the number densities of
each type, for systems brighter than M$_{\rm B} < -20$.
For example, 43\% of nearby bright barred galaxies are found in late
type disks, and late type disks with bars have a total number density of
5.2$\pm1.4$ $\times 10^{5}$ h$^{3}_{100}$ galaxies Gpc$^{-3}$. Not 
surprisingly,
we find that almost all (97\%) of barred galaxies are classified as
disks/S0s, with an
equal proportion belonging to early and late types.  Roughly 10\% of all
barred galaxies are S0s, while 3\% of classified 
barred galaxies are ellipticals and irregular galaxies.  The small
fraction of bars that are found in ellipticals shows that these
systems have been misclassified, as a bar needs a disk to exist.
  
By examining the fraction of
galaxies in each morphological type that are barred (plotted in Figure~6), 
we find that roughly 50\% of all disk galaxies are barred, while the fraction
of barred galaxies increases for later types.  There is also a fairly high
fraction of nearby galaxies with bars, with a number density in the nearby
universe of $\sim$1.2 $\times 10^6$ h$^{3}_{100}$ barred galaxies Gpc$^{-3}$.

\begin{figure}
\includegraphics[width=84mm]{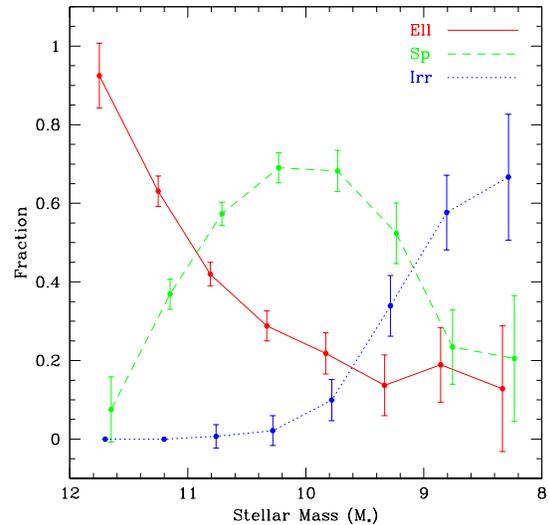}
 \caption{The fractional contribution of various morphological types to
the local  stellar mass density down to a lower limit of
M$_{*} = 10^{8}$ \solm. The solid line shows the ellipticals, the dashed
line spirals, and the dotted line irregular galaxies. }
 \label{sample-figure}
\end{figure}

\subsubsection{Rings}

Rings often follow the presence of bars in galaxies (Buta \& Combes 1996), 
and this can be quantitatively demonstrated as they occupy the same types of
galaxies.  Of all ringed galaxies brighter
than M$_{\rm B} < -20$, nearly all (98\%) are located in disks/S0s (Table~2).   
The fraction of ringed galaxies which are hosted inside ellipticals, S0s and 
irregulars is similar to the fraction with bars.  There are however
fewer ringed galaxies than barred galaxies in the nearby universe, with 
a number density of 3.8 $\times 10^{5}$ ringed galaxies Gpc$^{-3}$.  
Figure~6b shows the fraction of galaxies of all morphological types
with rings, revealing that early disks are more likely to host
a ring than later types.  

\setcounter{table}{2}
\begin{table*}
\begin{minipage}{190mm}
 \caption{Spearman Correlation Matrix For All Galaxies}
 \label{tab1}
 \begin{tabular}{@{}rrrrrrrrrrrrrrrr}
  \hline         
  & T-Type & R & $<\mu>_{e}$ & (U-B)$_{0}$ & (B-V)$_{0}$ & W$_{20}$ & V$_{\rm S}$ & HI & FIR & $\lambda$ & V$_{max}$ & M$_{B}$ & $I$ & $\rho$ & M$_{*}$ \\
  \hline
T-type &     1.00   &     -0.24        & {\bf  0.53}       &     {\bf  -0.79}        &     {\bf  -0.80}       &     {\bf  -0.51}       &     -0.43        & 0.18        & 0.13        & {\bf  0.91}        &     {\bf  -0.50}        & 0.25        & 0.23       &     -0.14       &     -0.68 \\
R        &     -0.24    &       1.00        & 0.08        & 0.23         & 0.20        & {\bf  0.66}        & 0.48       &     {\bf  -0.76}       &     {\bf  -0.55}       &     {\bf  -0.62}        & {\bf  0.65}       &     {\bf  -0.85}       &     -0.36       &     -0.13        & {\bf  0.51} \\
$<\mu>_{e}$  & {\bf  0.53}        & 0.08           & 1.00       &     -0.35       &     -0.36       &     -0.38        & 0.09       &     -0.01        & 0.08        & {\bf  0.56}       &     -0.35     &      0.00         & 0.07        &     -0.10       &     -0.07 \\
$(U-B)_{0}$  &     {\bf  -0.79}  & 0.23       &     -0.35           & 1.00        & {\bf  0.91}        & {\bf  0.51}         & {\bf  0.60}        & 0.04        & 0.03       &     {\bf  -0.67}        & 0.48       &     -0.23        &     -0.30        & 0.19        &     {\bf  0.75} \\
$(B-V)_{0}$  &     {\bf  -0.80}   &   0.20       &     -0.36        & {\bf  0.91}           & 1.00        & 0.47        & {\bf  0.57}        & 0.09        & 0.03       &     {\bf  -0.64}        & 0.47       &     -0.19       &     -0.28        & 0.22         & {\bf  0.80}  \\
W$_{20}$   &     {\bf  -0.51}        & {\bf  0.66}       &     -0.38        & {\bf  0.51}        & 0.47           & 1.00         & 0.30      &     {\bf  -0.54}       &     -0.39       &     {\bf  -0.72}        & {\bf  0.87}       &     {\bf  -0.68}       &     -0.38       &     -0.03   & {\bf  0.51} \\
V$_{\rm S}$       &     -0.43        & 0.48        & 0.09         & {\bf  0.60}    & {\bf  0.57}         & 0.30           & 1.00        & 0.09        & 0.07       &     -0.31        & 0.21       &     {\bf  -0.55}       &     -0.16        & 0.06        & {\bf  0.67} \\
HI   & 0.18       &     {\bf  -0.76}       &     -0.01        & 0.04        & 0.09       &     {\bf  -0.54}        & 0.09           & 1.00        & {\bf  0.56}        & {\bf  0.53}       &     {\bf  -0.55}        & {\bf  0.68}        & 0.23         & 0.2       &     -0.05 \\
FIR   & 0.13       &     {\bf  -0.55}        & 0.08        & 0.03        & 0.03       &     -0.39        & 0.07        & {\bf  0.56}           & 1.00         & 0.40       &     -0.46        & {\bf  0.64}        & 0.05        & 0.23       &     -0.08 \\
$\lambda$  & {\bf  0.91}       &     {\bf  -0.62}        & {\bf  0.56}       &     {\bf  -0.67}       &     {\bf  -0.64}       &     {\bf  -0.72}       &     -0.31        & {\bf  0.53}         & 0.40          & 1.00       &     {\bf  -0.72}       & {\bf  0.67}         & 0.40       &     -0.03   &     {\bf  -0.51} \\
V$_{\rm max}$    &     {\bf  -0.50}       & {\bf  0.65}       &     -0.35        & 0.48        & 0.47        & {\bf  0.87}        & 0.21       &     {\bf  -0.55}       &     -0.46       &     {\bf  -0.72}           & 1.00       &     {\bf  -0.69}       &     -0.36       &     {\bf  -0.05} &     {\bf  0.55} \\
M$_{\rm B}$  & 0.25       &     {\bf  -0.85}      &     0.00       &     -0.23       &     -0.19       &     {\bf  -0.68}       &     {\bf  -0.55}        & {\bf  0.68}        & {\bf  0.64}        & {\bf  0.67}       &     {\bf  -0.69}           & 1.00        & 0.31         & 0.10     &     {\bf  -0.59} \\
$I$        & 0.23       &     -0.36        & 0.07        &     -0.30       &     -0.28       &     -0.38       &     -0.16        & 0.23        & 0.05         & 0.40       &     -0.36        & 0.31           & 1.00       &     -0.01       &     -0.28 \\
$\rho$       &     -0.14       &     -0.13        &     -0.10        & 0.19        & 0.22       &     -0.03        & 0.06         & 0.20        & 0.23       &     -0.03       &     -0.05         & 0.10       &     -0.01           & 1.00    & 0.13 \\
M$_{*}$       &     {\bf  -0.68}        & {\bf  0.51}       &     -0.07        & {\bf  0.75}         & {\bf  0.80}        & {\bf  0.51}        & {\bf  0.67} &  -0.05       &     -0.08       &     {\bf  -0.51}   & {\bf  0.55}       &     {\bf  -0.59}       &     -0.28        & 0.13     & 1.00 \\
  \hline
 \end{tabular} \\
Spearman correlation matrix for all the galaxies in our sample.  Displayed
are the correlation between various physical properties, including
morphology (T-type), radius (R), surface brightness ($<\mu>_{e}$), colour
(both $(U-B)_{0}$ and $(B-V)_{0}$), HI line width (W$_{20}$), central
velocity dispersion (V$_{\rm S}$), HI and far infrared line fluxes (FIR), the
luminosity class ($\lambda$), the maximum rotation velocity (V$_{\rm max}$),
the absolute magnitude (M$_{\rm B}$), 
an interaction index  ($I$), local density ($\rho$) and the stellar mass
M$_{*}$.  The numbers in bold show the strongest correlations between
the various parameters.  Tables 4-6 show these correlations after
the same is divided into early/late, red/blue and bright/faint
systems.
\end{minipage}

\end{table*}

\subsubsection{Multiple Components}

In addition to bars and rings, the RC3 also includes a classification for
galaxies that have multiple components.  Intuitively, this might
correlate with the presence of merging activity. About
71\% of galaxies classified as having multiple components are
disks, while 11\%, 13\% and 5\% of multiple galaxies are ellipticals, S0s
and irregulars, respectively.  However, a different pattern emerges when
we consider the fractions of each galaxy type that have a multiple
classification.  Figure~6c plots the fraction of galaxies in each T-type
that have morphological evidence for multiple components.  This
demonstrates that only about 5 - 10\%
of all galaxies in the RC3 are classified as multiple galaxies.  
A higher fraction of irregulars are classified as having 
multiple components, although this is likely partially due to low number
statistics.  If these fractions are tracing the merging activity of galaxies 
in the nearby universe, the derived fraction of 5\% is close to the 
$z \sim 0$ merger fraction estimates from pairs (e.g., Patton et al. 2000). 

\section{Galaxy Properties and their Correlations}

\subsection{Statistical Analysis of the Nearby Galaxy Population}

We now examine how  the 14 properties described in \S 3.1 
correlate with each other, and how strongly.  We do this by performing 
Spearman 
rank correlation tests on various cuts of our data (Tables 3-6). 
These correlation analyses are performed on the entire sample (Table~3), when 
divided by early and late morphological types (Table~4), colour at 
$(B-V)_{0} = 0.7$ (Table~5), and absolute
magnitude at M$_{B} = -20$ (Table~6). Our sample for the Spearman rank
correlations include the RC3 sample out to radial velocities of 15,000 \kms.
We discuss in specific cases where not having a volume limited sample may
be affecting our results.

Our entire sample's correlation matrix (Table 3) reveals several strong 
relationships, including between: V$_{\rm S}$ and M$_{\rm B}$ (Faber-Jackson), 
log V$_{\rm max}$
and M$_{\rm B}$ (Tully-Fisher), and size (R) with M$_{\rm B}$ 
(Kormendy relation). However, the relations we find are weaker than
these scaling relationships, as the sample here has not been divided
into different morphological types.
In general however, there are many correlations between the
14 parameters, revealing that to one degree or another most 
properties of galaxies are interrelated.  Is this 
interrelation the result of a single principle, or a group of 
physical principles producing different observational effects?
For example, it is possible that size and M$_{\rm B}$ have a strong
correlation because they both fundamentally scale with an underlying feature,
which is likely the total mass? 
The same is also true for the strong correlation between both 
M$_{\rm B}$ and size
with internal velocities - as things get bigger and brighter they are
likely to have more mass, an effect traced by larger internal motions.

We divide these parameters into three classes, following the
ideas described in Bershady, Jangren \& Conselice 2000 (hereafter BJC00).
BJC00 determined from an analysis of nearby galaxy images that galaxies
can be classified according to scale, form, and spectral index.  In
this study colour represents the spectral index, scale includes the 
size, various fluxes, internal velocities and stellar mass, while form is 
given by the T-type, bars and the multiple component classification.    
We investigate which of these measurable properties 
are the most fundamental, and which are the most useful for
classifying galaxies.

\subsubsection{Form Correlations}

We use the T-type to represent the form of a galaxy and to first
order assume it correlates with 
the bulge to disk ratio of each galaxy (see also Conselice 2003).  T-types
(or Hubble types) have been known since at least Holmberg (1958) to correlate
broadly with other physical properties, such as recent star formation.  Almost
by definition this should be the case, as ellipticals and S0s, which populate
the earlier T-types, appear fundamentally different from later type galaxies.
This difference has been described in more detail by e.g.,  Roberts \& Haynes 
(1994).  It is not yet clear however which physical
properties of galaxies correlate strongest with morphology, and if Hubble 
morphology is enough to describe the most fundamental aspects of
galaxies.  With our Spearman correlation coefficients, we can begin to 
address this question.

{\small
\setcounter{table}{3}
\begin{table*}
\begin{minipage}{1000mm}
\vbox to70mm{\vfil
 \caption{Spearman Correlation Matrix For Early/(Late) Galaxies}
 \label{tab1}
 \begin{tabular}{@{}ccccccccccccccccc}
  \hline
  & T-Type & R & $<\mu>_{e}$ & $(U-B)_{0}$ & $(B-V)_{0}$ & W$_{20}$ & V$_{\rm S}$ & HI \\ &  FIR & $\lambda$ & V$_{\rm max}$ & M$_{B}$ & $I$ & $\rho$ & M$_{*}$ \\
  \hline
T-type & 1.00   (1.00)  &   -0.09   (-0.38)  &   -0.03   (0.51)  &   -0.38   (-0.61)  &   -0.42   (-0.63)  &   0.09   (-0.56)  &   -0.33   (-0.42)  &   -0.13   (0.29) \\  &   -0.01   (0.21)  &   0.26   (0.90)  &   0.05   (-0.55)  &   0.20   (0.41)  &   -0.05   (0.27)  &   -0.09   (-0.03)  &   -0.36   (-0.55)  \\   
\vspace{-0.23cm}
\\
R & -0.09   (-0.38)  &   1.00   (1.00)  &   0.41   (-0.09)  &   0.29   (0.41)  &   0.24   (0.36)  &   0.46   (0.68)  &   0.59   (0.24)  &   -0.54   (-0.78)  \\ &   -0.52   (-0.57) &   -0.27   (-0.64)  &   0.45   (0.68)  &   -0.84   (-0.86)  &   -0.37   (-0.35)  &   -0.14   (-0.14)  &   0.84   (0.81)  \\   
\vspace{-0.23cm}
\\
$<\mu>_{e}$   &  -0.03   (0.51)  &   0.41   (-0.09)  &   1.00   (1.00)  &   0.14   (-0.25)  &   0.11   (-0.27)  &   0.16   (-0.42)  &   0.21   (-0.14)  &   -0.34   (0.09)  \\  &   -0.16   (0.17)  &   -0.09   (0.57)  &   0.15   (-0.40)  &   -0.37   (0.22)  &   -0.10   (0.11)  &   -0.04   (-0.05)  &   0.44   (-0.21)  \\   
\vspace{-0.23cm}
\\
$(U-B)_{0}$ & -0.38   (-0.61)  &   0.29   (0.41)  &   0.14   (-0.25)  &   1.00  (1.00)  &   0.73   (0.83)  &   0.37   (0.56)  &   0.51   (0.61)  &   0.10   (-0.11) \\  &   0.26   (-0.07) &   -0.31   (-0.66)  &   0.30   (0.57)  &   -0.37   (-0.39)  &   -0.29   (-0.33)  &   0.18   (0.19)  &   0.55   (0.68)  \\\vspace{-0.23cm}
\\
$(B-V)_{0}$ & -0.42   (-0.63)  &   0.24   (0.36)  &   0.11   (-0.27)  &   0.73   (0.83)  &   1.00  (1.00)  &   0.32   (0.5)  &   0.47   (0.54)  &   0.09   (-0.05)  \\ &   0.26   (-0.06)  &   -0.33   (-0.63)  &   0.30   (0.54)  &   -0.29   (-0.34)  &   -0.24   (-0.31)  &   0.16   (0.18)  &   0.54   (0.71)  \\
\vspace{-0.23cm}
\\
W$_{20}$ & 0.09   (-0.56)  &   0.46   (0.68)  &   0.16   (-0.42)  &   0.37   (0.56)  &   0.32   (0.50)  &   1.00  (1.00)  &   0.29   (0.45)  &   -0.43   (-0.56) \\ &   -0.26   (-0.40)   &   -0.19   (-0.73)  &   0.89   (0.87)  &   -0.44   (-0.7)  &   -0.28   (-0.39)  &   -0.04   (-0.03)  &   0.41   (0.69)  \\   
\vspace{-0.23cm}
\\
V$_{\rm S}$ & -0.33   (-0.42)  &   0.59   (0.24)  &   0.21   (-0.14)  &   0.51   (0.61)  &   0.47   (0.54)  &   0.29   (0.45)  &   1.00  (1.00)  &   -0.09   (-0.03)  \\ &   0.01   (-0.08) &   -0.53   (-0.34)  &   0.2   (0.46)  &   -0.68   (-0.26)  &   -0.22   (-0.18)  &   0.01   (-0.03)  &   0.74   (0.55)  \\  
\vspace{-0.23cm}
\\
HI & -0.13   (0.29)  &   -0.54   (-0.78)  &   -0.34   (0.09)  &   0.1   (-0.11)  &   0.09   (-0.05)  &   -0.43   (-0.56)  &   -0.09   (-0.03)  &   1.00  (1.00)  \\ &   0.63   (0.56) &   0.27   (0.56)  &   -0.43   (-0.58)  &   0.51   (0.70)  &   0.2   (0.23)  &   0.21   (0.20)  &   -0.27   (-0.54)  \\ 
\vspace{-0.23cm}
\\
FIR & -0.01   (0.21)  &   -0.52   (-0.57)  &   -0.16   (0.17)  &   0.26   (-0.07)  &   0.26   (-0.06)  &   -0.26   (-0.40)  &   0.01   (-0.08)  &   0.63   (0.56) \\ &   1.00  (1.00)  &   0.02   (0.42)  &   -0.34   (-0.47)  &   0.58   (0.67)  &   0.12   (0.05)  &   0.32   (0.21)  &   -0.24   (-0.54)  \\
\vspace{-0.23cm}
\\
$\lambda$  & 0.26   (0.90)  &   -0.27   (-0.64)  &   -0.09   (0.57)  &   -0.31   (-0.66)  &   -0.33   (-0.63)  &   -0.19   (-0.73)  &   -0.53   (-0.34)  &   0.27   (0.56) \\ &   0.02   (0.42)  &   1.00  (1.00)  &   -0.18   (-0.73)  &   0.32   (0.69)  &   0.02   (0.41)  &   -0.28   (-0.02)  &   -0.48   (-0.73)  \\   
\vspace{-0.23cm}
\\
V$_{\rm max}$  & 0.05   (-0.55)  &   0.45   (0.68)  &   0.15   (-0.40)  &   0.3   (0.57)  &   0.30   (0.54)  &   0.89   (0.87)  &   0.20   (0.46)  &   -0.43   (-0.58) \\ &   -0.34   (-0.47)  &   -0.18   (-0.73)  &   1.00  (1.00)  &   -0.48   (-0.72)  &   -0.32   (-0.37)  &   -0.09   (-0.05)  &   0.39   (0.74)  \\
\vspace{-0.23cm}
\\
M$_{\rm B}$ & 0.20   (0.41)  &   -0.84   (-0.86)  &   -0.37   (0.22)  &   -0.37   (-0.39)  &   -0.29   (-0.34)  &   -0.44   (-0.70)  &   -0.68   (-0.26)  &   0.51   (0.70) \\ &   0.58   (0.67)  &   0.32   (0.69)  &   -0.48   (-0.72)  &   1.00  (1.00)  &   0.30   (0.31)  &   0.05   (0.12)  &   -0.94   (-0.88)  \\   
\vspace{-0.23cm}
\\
$I$ & -0.05   (0.27)  &   -0.37   (-0.35)  &   -0.1   (0.11)  &   -0.29   (-0.33)  &   -0.24   (-0.31)  &   -0.28   (-0.39)  &   -0.22   (-0.18)  &   0.20   (0.23) \\ &   0.12   (0.05)  &   0.02   (0.41)  &   -0.32   (-0.37)  &   0.30   (0.31)  &   1.00  (1.00)  &   0.00   (-0.01)  &   -0.22   (-0.35)  \\  
\vspace{-0.23cm}
\\
$\rho$ & -0.09   (-0.03)  &   -0.14   (-0.14)  &   -0.04   (-0.05)  &   0.18   (0.19)  &   0.16   (0.18)  &   -0.04   (-0.03)  &   0.01   (-0.03)  &   0.21   (0.20) \\ &   0.32   (0.21)  &   -0.28   (-0.02)  &   -0.09   (-0.05)  &   0.05   (0.12)  &   0.00   (-0.01)  &   1.00  (1.00)  &   0.08   (0.10)  \\
\vspace{-0.23cm}
\\
M$_{*}$     &      -0.36 (-0.55)   &    0.84 (0.81) &    0.44 (-0.21)    &  0.55 (0.68)    &   0.54 (0.71)  &   0.41 (0.69)   &   0.74 (0.55)  &  -0.27 (-0.54)  \\ &   -0.24 (-0.54)  &     -0.48 (-0.73)  &     0.39  (0.74)  &    -0.94 (-0.88)   &  -0.22 (-0.35)    &    0.08 (0.10)   &     1.00 (1.00) \\
  \hline
 \end{tabular}}
\end{minipage}
\end{table*}}

Table~3 shows the Spearman coefficients for the entire RC3 galaxy sample.
If we consider strong correlations those with coefficients $>|0.5|$ (listed
in bold) then
T-types correlate strongly with both $(B-V)_{0}$ and $(U-B)_{0}$ colours, 
W$_{20}$, the luminosity index ($\lambda$), V$_{\rm max}$, $\mu_{\rm m}$, 
and M$_{*}$.    This reveals several
things about what a T-type is measuring.  Since T-types correlate with B/T
ratios, then the strong correlation between T-types
and colour is not surprising, as it is well known that the relative 
contributions of bulge and disk light correlates with
old and young stellar populations. The correlation between the 
stellar masses of galaxies
and Hubble T-type (-0.68 coefficient), is stronger than the
correlation between T-type  and absolute magnitude (0.25), showing
that fundamentally the Hubble sequence (or bulge to disk
ratio) is partially one of stellar mass.  There is likewise
a strong correlation (coefficient 0.75-0.8) between stellar
mass and colour (Table~1), which we discuss in \S 5.1.2.

T-types also correlate strongly with other scale features, as the 
strong correlations between T-types and V$_{\rm max}$, 
W$_{20}$, V$_{\rm S}$ and
M$_{*}$ demonstrates.  The absolute magnitude and size of galaxies do not
show the same strong correlation, although later type galaxies are smaller and
fainter. This weaker correlation between T-type and magnitude/radius
might be the result of magnitude/radii being secondary scale features.  
This is because the measured size of a galaxy is influenced by both the form 
of a galaxy (i.e.,
how light is distributed, for example bulge vs. disk components) and the 
intrinsic scale, and magnitude is influenced by scale, and spectral type, as 
younger, bluer stars  will make a galaxy brighter. In this sense these
values are not clean measurements of the scale. However, the internal 
velocities
of galaxies and stellar masses are not affected by these secondary effects,
and they thus represent more accurately the scale of a galaxy.  

Table~4 shows the Spearman correlation matrix between our 14 
parameters when the sample has been divided into early 
(T $< 1$) and late type (T $> 1$) galaxies.  We repeat the
Spearman correlation analysis on both early and late types, to 
determine how correlations between T-types and the other parameters 
are affected by this division.  Table~4 shows
that for early-types galaxies with T $< 1$ there is very little to no 
correlation between T-type and other physical parameters, 
with the exception of perhaps colour. This is such that later-type
early-types (E6-7,S0) are bluer and have a slightly lower stellar mass.

When we examine correlations between T-types and physical properties
for the late-type galaxies
(Table~4) many of the correlations seen in Table~3 remain, but with
most weaker.  The correlation with colour for
the late types declines from $\sim -0.8$ to $-0.61$, as does
the correlation with stellar mass.  Interestingly, the mild correlation
between T-types and surface brightness ($\mu_{\rm e}$) seems
to remain for the late types, such that later-types have a lower
surface brightness. This is perhaps the only correlation in
the Hubble sequence (besides V$_{\rm max}$) which is driven
mostly by later type galaxies alone.

Most correlations between physical features and Hubble types
is strongest when considering the entire sequence. One exception
to this is the correlation between stellar mass and colour which
remains strong after dividing the sample into early/late Hubble types 
(Table~4).  This shows
that this correlation is not driven entirely by morphology. 
There is however little correlation within the early-types themselves, 
with the exception of scale-scale correlations.
This demonstrates that the division of early
types onto the Hubble sequence, which is typically done by estimating
the axis ratios of galaxies (Sandage 1961), is largely devoid of much
useful physical information (Kormendy \& Bender 1996). 
The fact that T-types correlate strongly with colour and stellar mass 
can be understood to first order if star 
formation follows the presence of disks. Galaxies with types T $> 1$ have a 
diversity in B/T ratios, and a mix of segregated stellar populations 
(Papovich et al. 2003),  while ellipticals have B/T $\sim 1$ and tend to have 
nearly homogenous old stellar populations. 

\subsubsection{Spectral Type Correlations}

As discussed in \S 5.1.1 galaxy T-types correlate strongly
with colour.  Spectral type, as denoted by  either the
$(B-V)_{0}$ or $(U-B)_{0}$ colour, also correlates well with the 
internal velocities of galaxies, i.e. W$_{20}$, V$_{\rm max}$, and V$_{\rm S}$ 
as well as stellar masses (Table~3).
The colours of galaxies however do not strongly correlate with any other
galaxy property, including sizes, HI/FIR luminosities and absolute magnitude.
  
Late-types are blue, and systems with lower internal
velocities, and lower stellar masses are also bluer, as discussed above. 
There is no a priori reason for galaxies with lower internal velocities, 
lower stellar masses, to have average younger stellar populations, 
while more massive systems have a larger fraction of older stars.  
This correlation between ongoing star formation, as revealed through
bluer colours, and stellar mass is a form of the downsizing seen
up to $z \sim 1$ (Bundy et al. 2005, 2006), which is still an unexplained
physical effect that dominates the way galaxies formed.  

{\small
\setcounter{table}{4}
\begin{table*}
\begin{minipage}{1000mm}
\vbox to70mm{\vfil
 \caption{Spearman Correlation Matrix For Red/(Blue) Galaxies Divided at $(B-V)_{0} = 0.7$}
 \label{tab1}
 \begin{tabular}{@{}ccccccccccccccccc}
  \hline
  & T-Type & R & $<\mu>_{e}$ & $(U-B)_{0}$ & $(B-V)_{0}$ & W$_{20}$ & V$_{\rm S}$ & HI \\ &  FIR & $\lambda$ & V$_{\rm max}$ & M$_{B}$ & $I$ & $\rho$ & M$_{*}$ \\
  \hline
T-type & 1.00   (1.00)  & 0.02   (-0.20)  &   0.07   (0.54) &  -0.58   (-0.43) &   -0.55   (-0.48) &   0.09   (-0.48) &   -0.38   (-0.27) &   -0.41   (0.19) \\&  -0.24   (0.23) &  0.72   (0.88) &   0.16  (-0.45) &  0.07   (0.28) &   -0.01   (0.21) &   0 .00 (0.00) &   -0.23   (-0.39) \\   
\vspace{-0.23cm}
\\
R & 0.02   (-0.20) &   1.00   (1.00) &  0.48   (0.02) &   0.06   (0.39) &   0.03   (0.34)  &   0.36   (0.61) &   0.56   (0.40)  &  -0.63   (-0.81) \\  &  -0.49   (-0.55) &  0.00   (-0.60) &  0.40 (0.62)  &   -0.88   (-0.87) &   -0.11   (-0.36) &   -0.04   (0.01) &   0.83   (0.84) \\   
\vspace{-0.23cm}
\\
$<\mu>_{e}$  & 0.07   (0.54)  &   0.48   (0.02) &   1.00   (1.00) &  -0.08   (-0.14) &   -0.05   (-0.20) &   -0.04   (-0.36) &   0.18   (0.10)  &  -0.40   (0.01)\\ &  -0.22   (0.21) &   0.23   (0.46) &   0.09   (-0.28) &   -0.43  (0.13) &   -0.01  (0.04) &   -0.08   (-0.02) &  0.38   (-0.16) \\   
\vspace{-0.23cm}
\\
$(U-B)_{0}$ & -0.58   (-0.43) &   0.06   (0.39) &   -0.08   (-0.14) &   1.00   1.00 &   0.70   (0.75) &   0.03   (0.48) &   0.52   (0.35) &  0.28   (-0.22) \\ &   0.39   (-0.19) &   -0.39   (-0.60) &   -0.04   (0.49) &   -0.14   (-0.37) &   -0.02   (-0.31) &   0.06   (0.09) &   0.32   (0.57) \\   
\vspace{-0.23cm}
\\
$(B-V)_{0}$ & -0.55   (-0.48) &   0.03   (0.34) &  -0.05   (-0.20) &   0.70   (0.75) &   1.00   (1.00) &   0.00   (0.44) &   0.48   (0.20) &  0.27  (-0.16) \\ &   0.35   (-0.16) &   -0.24   (-0.59) &   -0.05   (0.48) &   -0.08   (-0.33) &   0.06   (-0.28)  & 0.08   (0.08) &   0.35  (0.62) \\   
\vspace{-0.23cm}
\\
W$_{20}$ & 0.09   (-0.48) &   0.36   (0.61) &   -0.04   (-0.36) &   0.03   (0.48) &   0.00   (0.44) &   1.00   (1.00) &   0.25   (0.32) &  -0.32   (-0.55) \\ &   -0.20   (-0.41)  &   -0.05   (-0.66) &   0.8   (0.85) &   -0.34   (-0.66)  &   -0.21   (-0.43) &   0.00   (0.05) &   0.34   (0.69) \\   
\vspace{-0.23cm}
\\
V$_{\rm S}$ & -0.38   (-0.27) &   0.56   (0.40) &   0.18   (0.10) &   0.52   (0.35) &   0.48   (0.20) &   0.25   (0.32) &   1.00   (1.00) &   -0.03   (-0.30) \\ & 0.11   (-0.38) &   -0.20   (-0.36) &   0.20   (0.36) &   -0.66   (-0.39) &   0.00   (-0.24) &   0.08   (-0.17) &   0.72   (0.46) \\   
\vspace{-0.23cm}
\\
HI & -0.41   (0.19) &   -0.63   (-0.81) &   -0.4   (0.01) &   0.28   (-0.22) &   0.27   (-0.16) &   -0.32   (-0.55) &   -0.03   (-0.3) &   1.00   (1.00) \\ &  0.57   (0.53) &   -0.05   (0.52) &  -0.41  (-0.58) &  0.58   (0.78) &   0.04   (0.25) &   0.09   (0.10) &   -0.50   (-0.68) \\   
\vspace{-0.23cm}
\\
FIR  & -0.24   (0.23) & -0.49   (-0.55) &   -0.22   (0.21) &   0.39   (-0.19) &   0.35   (-0.16) &   -0.20   (-0.41) &   0.11   (-0.38) &   0.57   (0.53) \\  & 1.00   (1.00) &   -0.06   (0.45) &   -0.30   (-0.50) &   0.52   (0.73) &   0.01   (0.03)  &  0.03   (0.13)  & -0.43   (-0.66) \\   
\vspace{-0.23cm}
\\
$\lambda$  & 0.72   (0.88) &  0.00   (-0.60) &   0.23   (0.46)  &   -0.39   (-0.60) &   -0.24   (-0.59) &   -0.05   (-0.66) &   -0.2   (-0.36) &  -0.05   (0.52) \\  &  -0.06   (0.45) &   1.00   (1.00) &   -0.09   (-0.67) &   -0.01   (0.67) &   0.10   (0.37) &   -0.09   (0.00) &   -0.02   (-0.73) \\   
\vspace{-0.23cm}
\\
V$_{\rm max}$  & 0.16   (-0.45) &  0.40 (0.62) &  0.09   (-0.28)  &   -0.04   (0.49) &   -0.05   (0.48) &   0.80   (0.85) &   0.20    (0.36) &   -0.41   (-0.58) \\ &   -0.30   (-0.50) &   -0.09   (-0.67) &   1.00   (1.00) &   -0.44   (-0.70) &   -0.20   (-0.40) &   -0.09   (0.01) &   0.43   (0.74) \\   
\vspace{-0.23cm}
\\
M$_{\rm B}$ & 0.07   (0.28) &  -0.88   (-0.87) &  -0.43   (0.13) &   -0.14   (-0.37) &   -0.08   (-0.33) &   -0.34   (-0.66) &   -0.66   (-0.39) &   0.58   (0.78) \\  &   0.52   (0.73) &   -0.01   (0.67) &   -0.44   (-0.70) &   1.00   (1.00) &   0.11   (0.32) &   0.01   (0.04) &   -0.95   (-0.93) \\   
\vspace{-0.23cm}
\\
$I$  & -0.01   (0.21) &   -0.11   (-0.36) &   -0.01   (0.04) &   -0.02  (-0.31) &   0.06   (-0.28) &   -0.21   (-0.43) &   0.00   (-0.24) &  0.04  (0.25) \\ &   0.01   (0.03) &   0.10  (0.37) &  -0.20   (-0.40)  &  0.11   (0.32) &   1.00   (1.00) &   -0.02   (-0.03) &  -0.07   (-0.35) \\   
\vspace{-0.23cm}
\\
$\rho$ & 0.00   (0.00) &   -0.04   (0.01) &   -0.08   (-0.02) &   0.06   (0.09) &   0.08   (0.08) &   0.00   (0.05) &   0.08   (-0.17) &  0.09   (0.10) \\ &   0.03   (0.13) &   -0.09   (0.00) &   -0.09   (0.01) &   0.01   (0.04) &  -0.02  (-0.03) &   1.00   (1.00) &   0.02   (0.01)   \\
\vspace{-0.23cm}
\\
M$_{*}$   &   -0.23 (-0.39)  &      0.83 (0.84)  &     0.38 (-0.16)  &       0.32 (0.57)  &     0.35 (0.62) &  0.34 (0.69)  &   0.72 (0.46)  &    -0.50 (-0.68) \\ &    -0.43 (-0.66) &   -0.02 (-0.73)  &   0.43 (-0.74)  &    -0.95 (-0.93) &     -0.07  (-0.35) &   0.02 (0.01)   &       1.00 (1.00) \\
  \hline
 \end{tabular}}
\end{minipage}
\end{table*}}

We separated the RC3 sample into red and blue systems, divided
at $(B-V)_{0} = 0.7$, and then
redetermine how these various correlations change (Table~5). We find a 
similar effect
as when we divide the sample by morphology.  Red galaxies 
show fewer correlations between different parameters (Table~5), again
with the exception of colour, such that later-types with red colours,
are bluer.  However, as for early-types, there is little
correlation between colour and stellar mass for red galaxies. Star
forming blue galaxies have physical 
properties that correlate well with each other, i.e. stellar mass
with colour, HI with stellar mass, and size with V$_{\rm max}$.  Interestingly,
the correlation between T-types and colour within the red galaxies is
slightly stronger than it is for the blue galaxies 
($\sim -0.55$ vs. $-0.45$).  There is still a fairly strong ($\sim 0.5$) 
correlation between $(B-V)_{0}$ colour with
internal velocity dispersion (V$_{\rm S}$), and size  with V$_{\rm S}$
(0.56), for red galaxies, revealing that these relationships
are independent of star formation.  

\subsubsection{Scale}

As was mentioned earlier, the different scale features of galaxies, namely:
size, stellar mass, internal velocities, magnitude, HI and FIR magnitudes, 
all correlate with each other with coefficients $\sim 0.5 - 0.7$ for most 
correlations. 
Some of these strong correlations are:  R with: W$_{20}$ (0.66), V$_{\rm S}$
(0.48), V$_{\rm max}$ (0.65), M$_{\rm B}$ (-0.85), and M$_{*}$ (0.51), 
and M$_{\rm B}$ with: W$_{20}$ (-0.68), V$_{\rm S}$ (-0.55) and 
V$_{\rm max}$ (-0.69).   This indicates that measures 
of scale correlate with each other, and are independent of classification.
There is also a correlation
between scale parameters, particularly internal velocities, and 
the morphological T-type, as discussed in \S 5.1.1.

Correlations between scale parameters remain when examining
galaxies divided by morphology, colour and absolute magnitude 
(Tables 4 $-$ 6).  For example, the correlation between 
M$_{\rm B}$ and R remains quite strong 
no matter how the sample is divided with coefficients of $\sim0.85$.
Other strong correlations are between internal velocities, V$_{\rm S}$, 
for early-types, V$_{\rm max}$ for late-types and absolute
magnitude (M$_{\rm B}$), with values $-0.68, -0.72$, respectively.

Correlations for systems divided into blue and red galaxies at
(B$-$V)$=0.7$, and for systems divided into
`bright' galaxies with M$_{B} > -20$, and faint galaxies with M$_{B} < -20$ 
show a similar dichotomy as late/early types (Table~5-6). Likewise,
as for red and early-types, the bright systems tend to have weaker
correlations compared to when the entire magnitude range is considered.

{\small
\setcounter{table}{5}
\begin{table*}
\begin{minipage}{1000mm}
\vbox to70mm{\vfil
 \caption{Spearman Correlation Matrix For Bright/(Faint) Galaxies Divided at M$_{\rm B} = -20$}
 \label{tab1}
 \begin{tabular}{@{}ccccccccccccccccc}
  \hline
  & T-Type & R & $<\mu>_{e}$ & $(U-B)_{0}$ & $(B-V)_{0}$ & W$_{20}$ & V$_{\rm S}$ & HI \\ &  FIR & $\lambda$ & V$_{\rm max}$ & M$_{B}$ & $I$ & $\rho$ & M$_{*}$ \\
  \hline
T-type & 1.00  (1.00)  &   -0.08    (-0.16)  &   0.38   (0.68)  &   -0.80   (-0.75)  &   -0.81   (-0.76)  &   -0.24   (-0.51)  &   -0.52   (-0.19)  &   -0.17   (0.10) \\ &   0.11   (0.31)  &   0.67   (0.95)  &   -0.29   (-0.51)  &   0.02   (0.38)  &   0.06   (0.21)  &   -0.14   (...)  &   -0.65   (-0.64)  \\\vspace{-0.23cm}
\\
R & -0.08   (-0.16)  &   1.00  (1.00)  &   0.28   (-0.04)  &   0.15   (0.16)  &   0.13   (0.13)  &   0.44   (0.49)  &   0.29   (0.24)  &   -0.52   (-0.71)  \\ &   -0.29   (-0.42)  &   -0.24   (-0.41)  &   0.41   (0.50)  &   -0.66   (-0.73)  &   -0.20   (-0.31)  &   -0.13   (...)  &   0.54   (0.52)  \\   
\vspace{-0.23cm}
\\
$<\mu>_{e}$   & 0.38   (0.68)  &   0.28   (-0.04)  &   1.00  (1.00)  &   -0.17   (-0.54)  &   -0.21   (-0.55)  &   -0.17   (-0.49)  &   0.11   (-0.31)  &   -0.28   (-0.08) \\  &   0.22   (0.22)  &   0.28   (0.63)  &   -0.17   (-0.46)  &   -0.20   (0.32)  &   -0.04   (0.06)  &   -0.10   (...)  &   0.00   (-0.51)  \\   
\vspace{-0.23cm}
\\
$(U-B)_{0}$ & -0.8   (-0.75)  &   0.15   (0.16)  &   -0.17   (-0.54)  &   1.00  (1.00)  &   0.91   (0.90)  &   0.4   (0.47)  &   0.65   (0.54)  &   0.27   (0.19) \\  &   0.3   (-0.16)  &   -0.55   (-0.6)  &   0.37   (0.52)  &   -0.07   (-0.31)  &   -0.21   (-0.23)  &   0.19   (...)  &   0.74   (0.78)  \\ 
 \vspace{-0.23cm}
\\
$(B-V)_{0}$ & -0.81   (-0.76)  &   0.13   (0.13)  &   -0.21   (-0.55)  &   0.91   (0.90)  &   1.00  (1.00)  &   0.36   (0.44)  &   0.61   (0.53)  &   0.29   (0.25) \\ &   0.23   (-0.17)  &   -0.48   (-0.59)  &   0.38   (0.49)  &   -0.06   (-0.28)  &   -0.21   (-0.16)  &   0.22   (...)  &   0.80   (0.81)  \\   \vspace{-0.23cm}
\\
W$_{20}$ & -0.24   (-0.51)  &   0.44   (0.49)  &   -0.17   (-0.49)  &   0.40   (0.47)  &   0.36   (0.44)  &   1.00  (1.00)  &   0.23   (0.34)  &   -0.23   (-0.39) \\ &   -0.12   (-0.39)  &   -0.35   (-0.62)  &   0.78   (0.82)  &   -0.39   (-0.59)  &   -0.22   (-0.34)  &   -0.03   (...)  &   0.47   (0.67)  \\   \vspace{-0.23cm}
\\
V$_{\rm S}$ & -0.52   (-0.19)  &   0.29   (0.24)  &   0.11   (-0.31)  &   0.65   (0.54)  &   0.61   (0.53)  &   0.23   (0.34)  &   1.00  (1.00)  &   0.25   (0.10) \\  &   0.21   (0.14)  &   -0.34   (-0.32)  &   0.18   (0.16)  &   -0.34   (-0.38)  &   -0.13   (-0.19)  &   0.05   (...)  &   0.66   (0.57)  \\   
\vspace{-0.23cm}
\\
HI & -0.17   (0.10)  &   -0.52   (-0.71)  &   -0.28   (-0.08)  &   0.27   (0.19)  &   0.29   (0.25)  &   -0.23   (-0.39)  &   0.25   (0.10)  &   1.00  (1.00) \\ &   0.32   (0.42)  &   0.07   (0.26)  &   -0.25   (-0.43)  &   0.39   (0.55)  &   -0.02   (0.20)  &   0.21   (...)  &   -0.08   (-0.14)  \\   
\vspace{-0.23cm}
\\
FIR & 0.11   (0.31)  &   -0.29   (-0.42)  &   0.22   (0.22)  &   0.30   (-0.16)  &   0.23   (-0.17)  &   -0.12   (-0.39)  &   0.21   (0.14)  &   0.32   (0.42)  \\ &   1.00  (1.00)  &   0.20   (0.44)  &   -0.24   (-0.45)  &   0.43   (0.54)  &   -0.11   (0.05)  &   0.22   (...)  &   -0.12   (-0.12)  \\
\vspace{-0.23cm}
\\
$\lambda$  & 0.67   (0.95)  &   -0.24   (-0.41)  &   0.28   (0.63)  &   -0.55   (-0.60)  &   -0.48   (-0.59)  &   -0.35   (-0.62)  &   -0.34   (-0.32)  &   0.07   (0.26) \\ &   0.20   (0.44)  &   1.00  (1.00)  &   -0.37   (-0.64)  &   0.26   (0.63)  &   0.13   (0.32)  &   -0.02   (...)  &   -0.46   (-0.78)  \\ 
\vspace{-0.23cm}
\\ 
V$_{\rm max}$  & -0.29   (-0.51)  &   0.41   (0.50)  &   -0.17   (-0.46)  &   0.37   (0.52)  &   0.38   (0.49)  &   0.78   (0.82)  &   0.18   (0.16)  &   -0.25   (-0.43)  \\ &   -0.24   (-0.45)  &   -0.37   (-0.64)  &   1.00  (1.00)  &   -0.42   (-0.60)  &   -0.21   (-0.32)  &   -0.05   (...)  &   0.52   (0.67)  \\ 
  \vspace{-0.23cm}
\\
M$_{\rm B}$ & 0.02   (0.38)  &   -0.66   (-0.73)  &   -0.20   (0.32)  &   -0.07   (-0.31)  &   -0.06   (-0.28)  &   -0.39   (-0.59)  &   -0.34   (-0.38)  &   0.39   (0.55) \\  &   0.43   (0.54)  &   0.26   (0.63)  &   -0.42   (-0.60)  &   1.00  (1.00)  &   0.11   (0.28)  &   0.10   (...)  &   -0.61   (-0.75)  \\
\vspace{-0.23cm}
\\
$I$ & 0.06   (0.21)  &   -0.20   (-0.31)  &   -0.04   (0.06)  &   -0.21   (-0.23)  &   -0.21   (-0.16)  &   -0.22   (-0.34)  &   -0.13   (-0.19)  &   -0.02   (0.20) \\  &   -0.11   (0.05)  &   0.13   (0.32)  &   -0.21   (-0.32)  &   0.11   (0.28)  &   1.00  (1.00)  &   -0.01   (...)  &   -0.25   (-0.28)  \\
\vspace{-0.23cm}
\\
$\rho$ & -0.14   (...)  &   -0.13   (...)  &   -0.10   (...)  &   0.19   (...)  &   0.22   (...)  &   -0.03   (...)  &   0.05   (...)  &   0.21   (...) \\ &   0.22   (...)  &   -0.02   (...)  &   -0.05   (...)  &   0.10   (...)  &   -0.01   (...)  &   1.00  (...)  &   0.15   (...)  \\   
\vspace{-0.23cm}
\\
M$_{*}$ & -0.65   (-0.64)  &   0.54   (0.52)  &   0.00   (-0.51)  &   0.74   (0.78)  &   0.80   (0.81)  &   0.47   (0.67)  &   0.66   (0.57)  &   -0.08   (-0.14) \\ &   -0.12   (-0.12)  &   -0.46   (-0.78)  &   0.52   (0.67)  &   -0.61   (-0.75)  &   -0.25   (-0.28)  &   0.15   (...)  &   1.00  (1.00)  \\   
  \hline
 \end{tabular}}
\end{minipage}
\end{table*}}

For the more luminous galaxies,  parameters that correlate
strongly are: magnitude/stellar mass with size scalings
and internal velocity (V$_{\rm S}$), as well as 
M$_{*}$, colour and T-type among each other (Table~6).  
This is also the case for  fainter
galaxies where there is a strong correlation ($\sim 0.8$) between
M$_{*}$ and colour, such that more massive galaxies are
redder.  Both W$_{20}$ and V$_{\rm max}$ correlate
well with most parameters except with the $I$ index, an indication
that for fainter systems V$_{\rm max}$ is measuring something
fundamental about the scale of these systems.  This is also
true when examining the correlation of internal
velocities (V$_{\rm S}$) of the more massive systems.  

\subsubsection{Local Density}

The local density (hereafter environment) of a galaxy has been known since
Hubble (1926) to correlate with galaxy populations.  This morphology-density
relation can be summarised by stating that in locally dense environments a large 
fraction of galaxies are early-types (either S0s or ellipticals), while in
lower density environments spiral galaxies are more common 
(e.g., Dressler 1980; Postman \& Geller 1984). 

We can reexamine the correlation of environment with other physical properties
by using the calculated galaxy number density ($\rho$) parameter (\S 3.1).
We however do not find a strong
correlation between $\rho$ and other parameters, including morphological
type.  This is perhaps because galaxies
of all types are found in all environments, but their relative proportions
can change, creating for example, the morphology-density
relation.   Table~3 shows that the correlations between
local density and properties such as colour and T-type have the general
trend expected, such as later types and bluer galaxies are found
in lower density environments.   However, 
these correlations are not as strong as many others.

\subsubsection{Summary of Correlations}

It is clear from the above discussion that there are strong correlations
between various galaxies properties.  A few of these correlations, 
such as size and brightness are very strong, and are independent of 
other properties such as recent star formation.  Perhaps the most
interesting correlation is the one between T-type, stellar
mass and colour, which all seem to relate to each other.

Scale-features, such as size and stellar mass correlate
strongly, which does not change after
dividing the population into early/late, red/blue or bright/faint
systems. However, other correlations break down when we examine
them within these subsets, showing the scale is a fundamental
feature of galaxies.

In particular, many correlations remain for late-type, or star forming
galaxies, as defined by (B$-$V) $< 0.7$, but do not appear as strong as
for the early/red galaxies.  These correlations, which break down in
the absence of recent star formation are: stellar mass/magnitude
with colour and HI/FIR magnitudes, and T-type with 
surface brightness.  There is also little 
correlation between T-types and size when galaxies are
broken into red/blue systems

Ongoing star formation is
therefore a critical component for trying to understand the properties
of galaxies.  Finally, if we accept the idea
that the strongest correlations with Hubble type reveal its  
underlying meaning, then the fact that T-types
correlate with colour and internal velocities demonstrates
that the Hubble sequence is fundamentally one of decreased total mass
and increased dominance of young stars, which accounts for its 
success in the nearby universe.

Note, that the form index, $I$, which has been shown to correlate
with dynamical disturbances (Conselice et al. 2000b), does not correlate 
strongly with any of
the other observational properties we study in this paper (Table~3).  
This reveals that $I$ is not directly related to, or
always causes, other parameters to change, with the exception of
perhaps the luminosity index. 

\subsection{Principal Component Analysis}

We carry out a principal component analysis of the properties of
our sample to determine the most fundamental features needed to
describe nearby galaxies.  The question we address is whether we can use 
only morphological type to describe a galaxy's major
properties.  We perform several principal component analyses 
(PCA) to determine which features cannot be accounted 
for by other properties, and to reduce the dimensionality of
our data set with its 14 parameters.  

PCA methods are commonly
used in astronomy, and good introductions to the topic 
can be found in e.g., Faber (1973), Whitmore
(1984), Han (1995), Brotherton (1996) and Madgwick et al. (2003).  
The basic goal of a PCA analysis is to reduce
the number of variables $(n)$ to a set of $m<n$ parameter combinations.
This is done through the calculation of eigenvectors
and eigenvalues for the different parameter combinations. The first
eigenvector is a line through the cloud of $n$ data points
which minimises the variance.  Remaining eigenvectors are
calculated until all the variance is explained.  Eigenvalues
reveal how much of the variation is accounted for by each
eigenvector.  Typically
the eigenvalue for an eigenvector must be $\geq 1$ to be significant.
Before we carry out our various PCA analyses, we linearise all
variables, subtract the mean from each variable, and normalise by
the standard deviation.

\setcounter{table}{6}
\begin{table}
 \caption{PCA Eigenvector (EV) Projections onto Galaxy Properties}
 \label{tab1}
 \begin{tabular}{@{}crrrrr}
  \hline
Property & EV 1 & EV 2 & EV 3 & EV 4 & EV 5 \\
\hline
T-type &   -0.06 & 0.61 & 0.03 & -0.07 & 0.12 \\
R      &   0.49 & 0.07 & 0.08 & 0.11 & -0.07 \\
$<\mu>_{e}$ & 0.10 & 0.42 & 0.36 & 0.35 & 0.03 \\
$(B-V)_{0}$  & 0.19 & -0.55 & 0.15 & 0.14 & -0.10 \\
Vel     &  0.19 & -0.12 & 0.07 & -0.34 & 0.90 \\   
HI   &  0.42 & 0.28 & -0.12 & -0.06 & -0.06 \\
FIR   & 0.14 & 0.06 & -0.69 & -0.41 & -0.19 \\
M$_{\rm B}$  & 0.50 & 0.09 & -0.06 & 0.02 & -0.04 \\
$I$        & -0.06 & -0.02 & -0.58 & 0.74 & 0.34 \\
M$_{*}$  & 0.47 & -0.20 & 0.05 & 0.11 & -0.05 \\
  \hline
Eigenvalues & 3.4 & 2.0 & 1.1 & 0.94 & 0.90 \\
 & 33.6\% & 19.7\% & 11.1\% & 9.4\% & 9.0\% \\
\hline
 \end{tabular}
\end{table}

We preformed a large number of PCA analyses on our data.  A typical analysis
that includes most of our properties is shown in Table~7.  We have created
a new index here called ``Vel'' which is the internal velocity of our
galaxies, V$_{\rm max}$, or V$_{\rm S}$ when V$_{\rm max}$ is not available. 
 When we calculate the total masses of 
our galaxies using $\sim R \times V_{\rm max}^{2}$, or total masses 
derived based on
scalings derived from models (Conselice et al. 2005b), we find the 
same correlations. 
The eigenvectors
listed in Table~7 show that several vectors are needed to reach a 
reasonable accounting of the total variance in galaxy properties.  The
total summation of the five eigenvectors listed in Table~7 represent
83\% of the variance in the properties of galaxies.  Due to the errors
on the measurements, this is likely the limit for meaningful eigenvectors.
In fact, when we plot the eigenvalues for the analysis in Table~7 vs.
principal component rank we find a shallow slope past EV-5 than before 
it, suggesting that these higher components are dominated by noise.

The first eigenvector (EV 1; Table~7) is dominated by 
M$_{\rm B}$, M$_{*}$, size (R), and HI content, which all
positively correlate (M$_{\rm B}$ was converted into a positive linear
flux for this analysis).  The first vector can therefore be thought of as
the scale of the system.  The second vector is dominated by the T-type,
the $(B-V)_{0}$ colour, and surface brightness.  The strong correlation between
Hubble types and colour suggests that this vector measures the presence
of recent star formation in a galaxy.  Vector 3 and Vector 4 are both 
dominated by the merger parameter ($I$) and the FIR luminosity, and as such 
it represents the merging properties of galaxies.  Vector 5 is dominated by
the internal velocities of galaxies. 

We find a similar pattern after we remove various variables from
the PCA analysis.  If we do not consider FIR or surface brightness 
in the analysis, we find that the eigenvectors are always 
dominated by $I$, $(B-V)_{0}$,
and M$_{*}$ in the first three principal components. We find in general
that the T-type is a significant contributor to many components after
the analysis is carried out through different cuts in the data.

When we do a basic PCA analysis on our sample separated into only T-types, 
a spectral
index ($(B-V)_{0}$ colour), the merger index ($I$), and stellar
mass (M$_{*}$) we obtain Table~8.  We find that the 
T-type, stellar mass and colour are highly correlated and dominate
the first component. The $I$ merger index defines the second 
principal component, while the third component is dominated
by the T-type, and the stellar mass.  This shows that the T-type is
successful as a general galaxy classification criteria as it correlates
strongly with colour and stellar mass, and several other properties 
(cf. Table~3). Therefore the three most fundamental properties for describing
nearby galaxies appear to be their scale, recent star formation, and the
presence of mergers.

\setcounter{table}{7}
\begin{table}
 \caption{EVs for PCA with T-type, Colour and M$_{*}$}
 \label{tab1}
 \begin{tabular}{@{}cccrrr}
  \hline
Property & EV 1 & EV 2 & EV 3 \\
\hline
T-type &   -0.57 & -0.13 & 0.60 \\
$(B-V)_{0}$ & 0.63 & -0.04 & -0.10 \\
$I$ & 0.06 & 0.96 & 0.23 \\
M$_{*}$  & 0.52 & -0.20 & 0.76 \\
  \hline
Eigenvalues & 2.0 & 1.02 & 0.64 \\
 & 50.8\% & 25.4\% & 16.1\%  \\
\hline
 \end{tabular}
\end{table}

From \S 5.1 and Table~3 we know that the interaction/merger index, $I$, does
not correlate strongly with any other galaxy property, with the exception of 
the luminosity index. The meaning of this is that galaxies which
are undergoing a merger cannot be accounted for by other measurable features. 
When we analyse only  galaxies undergoing mergers, defined as 
objects with $I > 1.5$, we find that three principal components are necessary 
to describe these systems.  In fact, for systems with $M > 1.5$ the 
correlation between T-type and $(B-V)_{0}$ colours breaks down.   
We conclude from the above analyses that our hypothesis, described in
Conselice (2003) that mass, star-formation and the presence of mergers are 
fundamental criteria, is a reasonable assumption.

\subsection{Galaxies in PCA Space}

In this section we examine whether the principal component projections
for our sample are correlated, and if different galaxy types separate
cleanly within PCA space (see Madgwick et al. 2003 for examples of
this using galaxy spectra).  For this analysis we examine the principal
components shown in Table~8.  Figure~9 shows the correlation between
the projections of our galaxies onto their first three eigenvectors listed
in Table~8, which
as we have discussed, generally correlate with colour (p$_{1}$), 
interactions/mergers (p$_{2}$), and stellar mass (p$_{3}$).  We furthermore
label the morphological type of galaxies on Figure~9, which shows that this 
relatively
simple PCA analysis separates nicely the major galaxy classes: early-types, 
disks, irregulars and merging galaxies.  The left panel of
Figure~9 shows a correlation between the projection of the eigenvectors
dominated by recent star formation (colour) and scale (stellar mass).  There
is however a perpendicular relation as well which is occupied mostly
by disk galaxies.

\begin{figure*}
 \vbox to 110mm{
\includegraphics[angle=0, width=144mm]{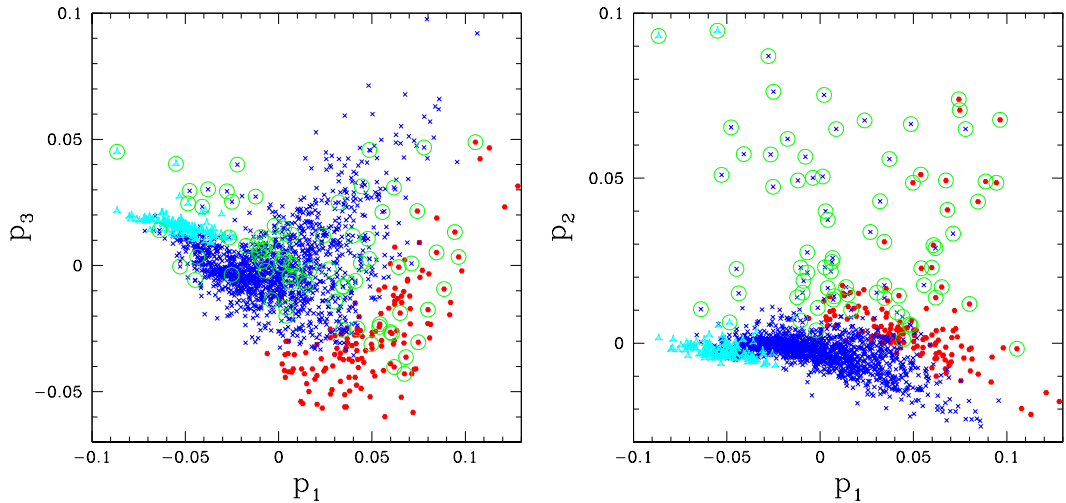}
 \caption{The relation between the projection of galaxies onto
the eigenvectors listed in Table~8.  The morphological and dynamical
state of each galaxies is also noted. The red solid points are the 
ellipticals and S0s, the blue crosses are the disks, and the cyan
triangles are irregulars.  The objects circled and in green are those
with an interaction/merger index $I > 2$.  The three projected
components, in which normal galaxies of different types, as well as
interacting/merging galaxies, are well separated, trace the scale (p$_{3}$),
star formation (p$_{1}$), and interacting/merging properties (p$_{2}$)
of galaxies (\S 5.2).}
} \label{fig}
\vspace{-1.5cm}
\end{figure*}

The correlation between p$_{1}$ and p$_{2}$ shows a similar patten, but
with only a slight correlation between the two eigenvector projections.  
However, it is clear that interacting
and merging galaxies, with $I > 2$, deviate from the nominal 
relation.  Examples of these high-$I$ systems are shown in Figure~2.
There is also a correlation between the second and third eigenvector
projections for most galaxies in our sample. Just as in the
right panel of Figure~9,  galaxies with a
high $I$ index deviate from this relation with a larger projection onto
the second eigenvector.

\subsection{The Strongest Correlations}

With our current data set we can address which scaling
relations are statistical the strongest for nearby galaxies.  While most 
of these relations are well known and characterised, we discuss several
below which are found independent of sample 
selection. These correlations are:  T-type with
$(B-V)_{0}$ colours, T-type with log V$_{\rm max}$, radius with absolute magnitude, radius with
V$_{\rm max}$, $(B-V)_{0}$ with log V$_{\rm max}$, and absolute magnitude with
log V$_{\rm max}$.  The relationship between these parameters are shown
in Figure~10.  The best fits between these parameters are,

$$ {\rm T} = 0.66\pm0.02 \times (B-V)_{0}^{-0.15\pm0.02},$$

$$ {\rm T} = 2.30\pm0.01 \times {\rm (log~V_{\rm max})}^{-0.06\pm0.003},$$

$$ {\rm M}_{\rm B} = -15.5\pm0.02 \times R^{0.1\pm0.004},$$

$$ {\rm log~V_{\rm max}} = 1.57\pm0.001 \times R^{0.12\pm0.01},$$

$${\rm log~V_{\rm max}} = 2.64\pm0.03 \times (B-V)_{0}^{0.16\pm0.01},$$

$$ {\rm M}_{\rm B} = -4.7\pm0.1 \times {\rm log~V_{\rm max}} - 9.8\pm0.2.$$

\begin{figure*}
 \vbox to 130mm{
\includegraphics[angle=0, width=134mm]{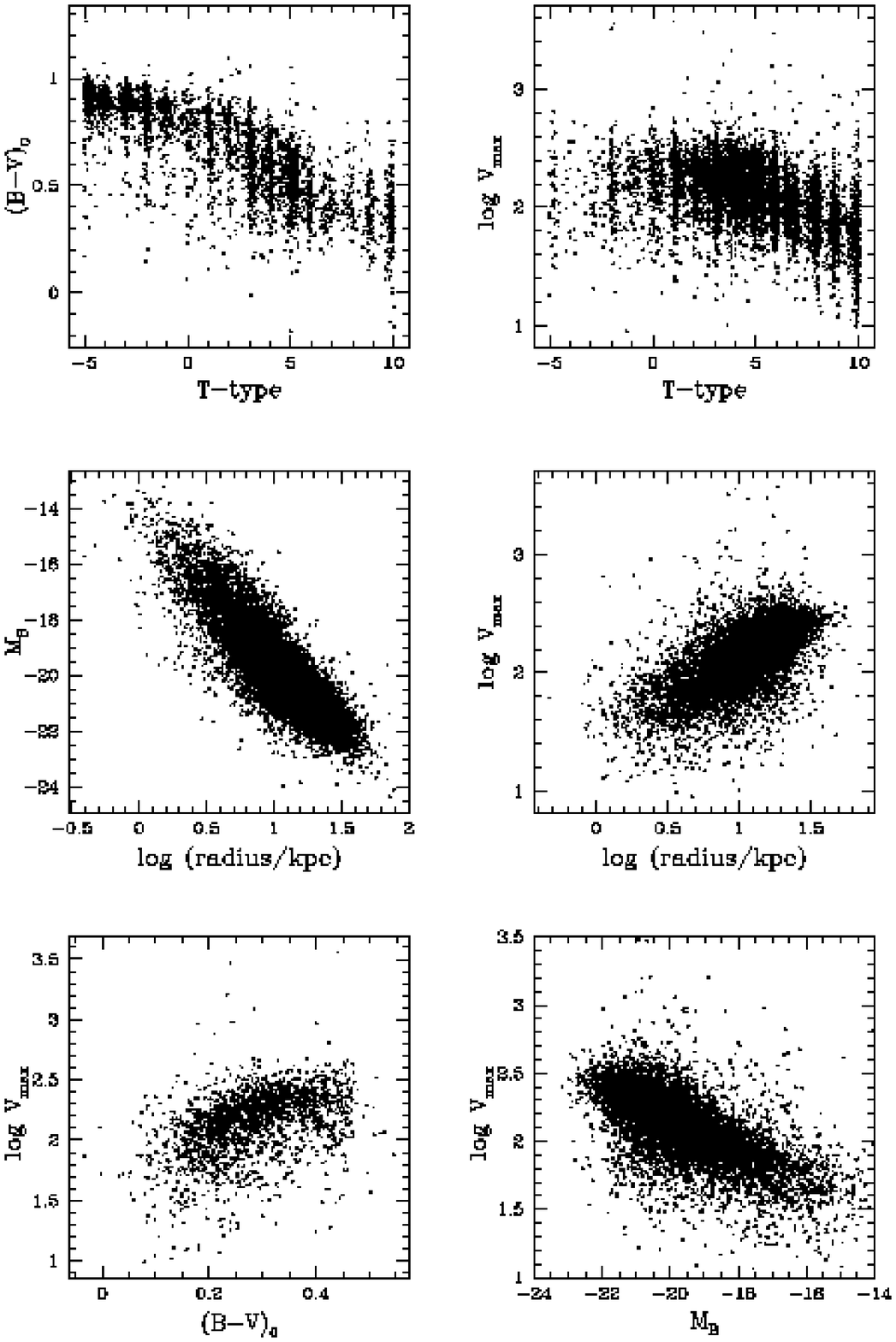}
 \caption{The strongest correlations found between the properties of the 
22,000+ galaxies studied in this paper. Most of these relationships are 
already well known, and described for morphological subsets, such as the 
Tully-Fisher relation, and the size-magnitude relation.  For our purposes, 
the most interesting trends are those of T-type with colour and 
V$_{\rm max}$, where considerable scatter can be seen.  This is also 
true for the correlation between colour and V$_{\rm max}$. The scatter 
in these correlations is larger than the observational errors, and 
suggests that while morphology, colour, and scale all correlate in 
nearby galaxies, they are still independent features.}
} \label{sample-figure}
\vspace{14cm}
\end{figure*}

\noindent Most of these scalings are already well known, and have been 
described
in various contexts previously.  Many become stronger when only 
considering certain types of galaxies (e.g., ellipticals for the 
Faber-Jackson, disks for Tully-Fisher).  We are particularly 
interested here in the correlation
of T-type with colour and V$_{\rm max}$ (as a proxy for other scale
features).  Figure~10 plots the relationship between these properties.
As can be seen, later T-types are both bluer and have lower internal
velocities.  These are in particlar strong correlations, as shown in
Table~3.  As we argue in this paper, these two correlations are
the reason why the Hubble sequence has been generally successful at
separating galaxy types.  However, as can be seen there is a large
scatter in these correlations, such that at every T-type there
is a wide range in colour and V$_{\rm max}$ values.  This is
also true for the relationship between $(B-V)_{0}$ and V$_{\rm max}$.
This diversity cannot be accounted for by errors on classifications, or 
on the measured values.  This shows that there are variations in colour and 
scale which is not accounted for by Hubble type.


\section{Discussion}

\subsection{Physical Classification}

For the remainder of this paper we address whether the results of the
first part of this paper can be used to construct a classification
system for galaxies that is physically meaningful.  First, 
it is worth asking if we 
need a physical classification system for galaxies.  Is it possible
that relations between observables and (non-morphological)
galaxy properties are enough to characterise the population? 
The first classification schemes by Wolf (1908), Hubble (1926),
de Vaucouleurs (1959) and Sandage (1961,1975) are purely descriptive, and
the idea that a galaxy's classification should be based solely on the most
obvious features seen in optical light persists (Sandage 2005).  The
reason given for this philosophy is that no physical understanding of
what drives galaxy formation/evolution should be used to create
a classification system.  As Sandage
(2005) explains, ``A good classification can drive the physics, but the 
physics must not be used to drive the classification.''

Sandage's (2005) argument, which follows Hubble's original philosophy, 
likely resulted from the
earliest attempts to classify galaxies through the use of theoretical arguments
by Jeans (1919).  Hubble quite rightly avoided theory when developing his
classification system, yet physical morphology does not necessarily rest
upon theory, but empirical observations.  Physical morphology in this
sense is not a theoretical morphology.  
If we knew with reasonable certainty 
what the physical effects were that drove the properties of 
galaxies, a classification based on these features would
be superior to theoretically, or subjectively based 
systems.

Several features of galaxies also  
make them fundamentally different from other objects 
which have been successfully classified based solely on their
appearance, including stars, plants/animals and
the chemical elements.  Galaxies are fundamentally different from
these types of objects as they are not stable or passive
systems, nor do they have a well defined life/death as do living organisms and 
stars.  Furthermore, elements, plants/animals and most stars spend most of
their life span in the same form. This is not the case for
galaxies, where it has been shown that systems of the same mass change
their appearance in optical light dramatically in the last 10 Gyr
(e.g., van den Bergh et al. 2001; Conselice et al. 2005a; 
Ravindrinath et al. 2006).  
Galaxies are not static systems - they are always changing and evolving
due to the complex mixtures of stars, gas, dust, dark matter, and from
interactions with other galaxies.  

Another reason that the physical
properties of galaxies can be used to devise a classification system
is the fact that it is not a given that the rest-frame optical light
of galaxies reveal the most important features of galaxies.  
For example, we now know that the bulk emission from 
an average galaxy is in the mid infrared (e.g., Dale et al. 2005), 
and if any wavelength
should be used to classify galaxies it would naturally be where most
of the light is emitted. Fundamentally, if we want to classify galaxies
through structure, the best approach is using a map of the distribution
of mass, which would be mostly dark matter.  A mapping such as this, with 
enough resolution for classification, does not currently exist.  In 
blue light, where the Hubble system was and is being used,
we are examining only the stars in galaxies, and often only
the brightest ones, which tend to be young.
This leads to physical classifications, where if we think we understand
the major properties of galaxies, we can utilise
these for classification.  

Knowing which properties of galaxies are not important for
classification is just as critical as uncovering the dominant ones.  
We can understand why through analogies with
other sciences.   In biological
taxonomy, such as separating different populations
of mammals, features such as hair or eye colour, or exact size, are not 
classification criteria.
Similarly, features such as rings, bars and even spiral arms
are important for the immediate appearance of a galaxy, but these are not 
basic features.  Classifications using these criteria
are purely descriptive. Their inclusion as major classification
criteria are also arbitrary, as other properties, such as the
steepness of light profiles, the clumpy nature of light seen in
many galaxies, the spokes and spurs seen in spiral arms, etc. were never 
considered important, or were technically infeasible to measure, 
by early morphologists.

We can use the results of this paper to construct a physically
based method for classifying galaxies. 
Physical morphology is not a new idea, and attempts to construct a meaningful
system for galaxies started with the work of Morgan (1958; 1959) who 
attempted to correlate ``the forms of certain galaxies and their stellar 
content as estimated from composite spectra'' from Morgan \& 
Mayall (1957).  Later attempts include van den Bergh's (1960) study
on the correlation between spiral arm structure and intrinsic brightness.
Modern studies have attempted to classify ellipticals by their structures
(Kormendy \& Bender 1996), and at using interacting and star formation
properties to classify all galaxies (e.g., Conselice 1997; Conselice et al.
2000a; Conselice et al. 2000b; Bershady et al. 2000).
As we argued in \S 5, star formation, interactions/mergers, and 
the masses of galaxies are the three most fundamental parameters that can be 
used together to form a new physical galaxy classification.

\subsubsection{Mass}

We concluded in \S 5 that the underlying
mass (scale) of a galaxy is a fundamental property, which is responsible
for the major component in our PCA analyses.  While few would argue
that the mass of a galaxy is not a fundamental criterion for classification,
it is not obvious that it is possible to classify galaxies based on their
mass, without direct measurements.

However, there are structural indicators in optical light which
can be used as a proxy for the underlying mass in a galaxy.
As shown by e.g., Graham et al. (2001) and Conselice (2003) the 
concentration of a galaxy's light, measured through either parametrised fits
to light distributions, or from direct measurements using a concentration
index (Bershady et al. 2000), broadly correlate with the stellar and kinematic
masses of galaxies.  This is true at low redshift, as well
as out to $z \sim 1$ (Conselice et al. 2005a). 

\begin{figure*}
 \vbox to 120mm{
\includegraphics[angle=0, width=154mm]{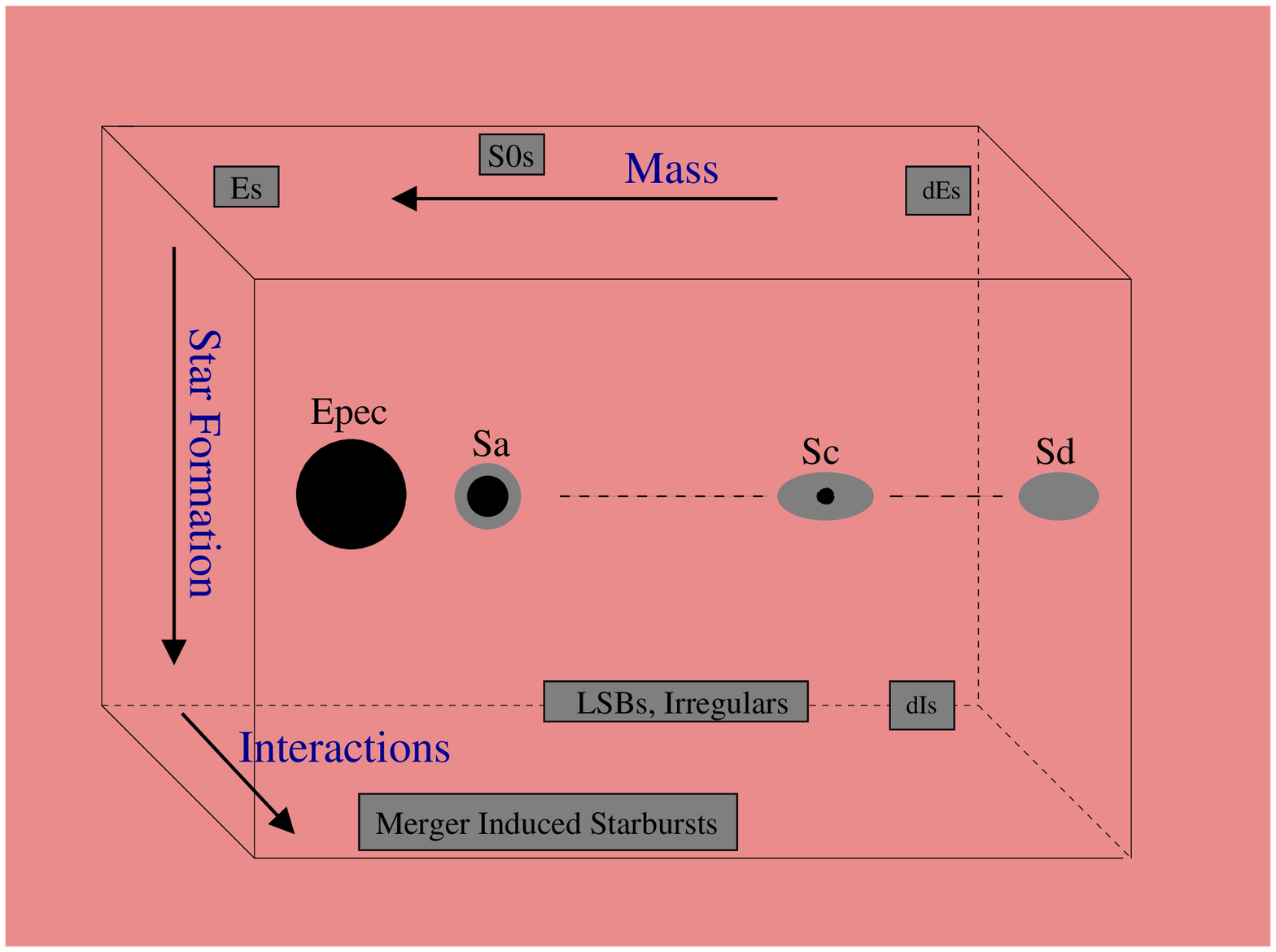}
 \caption{The three dimensional classification volume proposed in this paper 
to account for all known galaxies.  The three parameters used are the 
mass, recent star formation and interaction properties of galaxies.   
Several well known galaxy types are labelled in their appropriate locations in 
this space.  At the top of the classification cube are found galaxies with
no recent star formation.  Across the mass axis we progress from dEs
to S0s to giant ellipticals. Any galaxies with a recent interaction/merger
would be found further into the third dimension.  As we go further into
the cube within the y-axis, we find systems with a higher contribution of
recent star formation to the stellar mass. Low-surface brightness disks
would be found in a region of this space where the star formation is modestly
high, and interaction history low.  Merger induced starbursts are located
in the high star formation, high interaction part of the diagram, while
irregulars and dIs are located towards the bottom right of the cube, as
these are typically lower mass star forming galaxies.  The Hubble sequence is
a one dimension projection of the star formation and interaction axes 
onto the mass sequence.  This classification necessarily replaces 
idealised Platonic Hubble types by idealised parameters that are not always
easy to retrieve, but in principle are measurable for all observable
galaxies.}
} \label{fig}
\vspace{4cm}
\end{figure*}

The degree of apparent concentration of stellar light in 
a galaxy has been used (even if not explictly stated) since Hubble (1926) as 
the main axis for nearly all galaxy classification systems.  As shown 
in e.g., Conselice (2003) galaxy concentration relates to the bulge to
disk ratio and Hubble type, although this breaks down for dwarfs.
Furthermore, light concentration has continued 
as the main axis in all galaxy classification systems, such as in de 
Vaucouleurs (1959) and
Morgan (1959), and continues as a quantitative system through fitting bulge to
disk ratios.  

While bright Hubble types generally correlate with the masses of galaxies, they
do not account for other galaxy properties needed for a full classification.
The fact that late-type galaxies, which generally have low: masses, bulge to
disk ratios, and light concentrations, are blue is
a result of evolutionary processes, rather than an inherent success of
the classification. It is a coincidence that Hubble types
correlate well with colour in the nearby universe, as this is
not the case at higher redshifts where morphologically classified
ellipticals are often blue in colour and star forming (e.g., Stanford
et al. 2004; Bundy et al. 2005).  This correlation breaks down
even more at $z > 1$, where the Hubble sequence itself begins 
to dissipate.

\subsubsection{Star Formation}

As shown in \S 5 we find a morphological dichotomy between galaxies that 
are dominated by recent star formation and those that are not.  
Our principal component analyses reveals that colour is the major
contribution to a significant principal component, which is not accounted 
for solely by the Hubble type,
as galaxies at a given morphology type have a range of colours (\S 5.1,
Figure~10). Therefore, if we  knew the current star formation rate of 
a galaxy it would be a very powerful classification parameter.
Typically the current star formation rate is derived from either 
ultraviolet, H$\alpha$ or far-infrared fluxes (Kennicutt 1998), although
these measurements are typically time-consuming and hard to obtain
for large samples.   It is therefore desirable to use 
a method that can estimate the star formation properties of galaxies 
based on optical imaging.  For most galaxies, this can be done
by combining colour information along with high-frequency structural
features produced by young stars (Isserstedt \& Schindler 1986; 
Conselice et al. 2000a; Conselice 2003).  The clumpiness index, $S$,
(Conselice 2003) is a good measure of the unobscurbed current star formation 
rate in galaxies, and can therefore be used as an indicator.

\subsubsection{Galaxy Mergers/Interactions}

Like the scale of a galaxy and its current star formation rate, galaxy
mergers and interactions are an important aspect for driving galaxy evolution.
Although mergers are rarely found in the local galaxy population, they have 
strong effects on the evolution of galaxies and their properties.   
Ellipticals and high surface brightness bulges are likely the 
result of major mergers, and other accretion processes.
In reality, all merger events for any one galaxy are nearly impossible to
trace with certainty, especially old mergers that have
gone through violent relaxation, after which the original orbital 
information is lost. 
Recent interactions between galaxies however will almost always have obvious 
signatures that will change the way galaxies appear, as well as their 
internal evolution, as has been shown in various methodologies.

\subsection{A New Classification System: The Classification Cube}
 
Figure~11 shows how
galaxies of various types fit into a 3-dimensional parameter space defined
by galaxy mass, recent star formation, and interactions.
On this diagram, the $x$-axis signifies the mass, while
the $y$-axis represents the recent star formation, and the
$z$-axis represents the degree of recent interactions with other galaxies.

As opposed to previous classification systems, the one presented in 
Figure~11 can account for all the major galaxy types described in \S 2.
Classical ellipticals, 
with little sign of recent interactions or star formation will occupy
the high-mass, low star formation and low interaction corner.   The
top of this classification volume also contains the S0s, dwarf 
ellipticals/spheroidals. S0s have little evidence for recent star formation,
but are lower-mass systems than ellipticals, and they occupy the 
same star formation and interaction space as the ellipticals.  dEs are 
systems with low-mass, and as such they occupy the far
end of the mass sequence for these galaxies.  The dwarf irregulars have
a similar internal structure and mass to the dEs, except they contain star 
formation, and thus fall in the same mass and interaction space, but are
separated in star formation space.

\begin{figure}
\includegraphics[width=84mm]{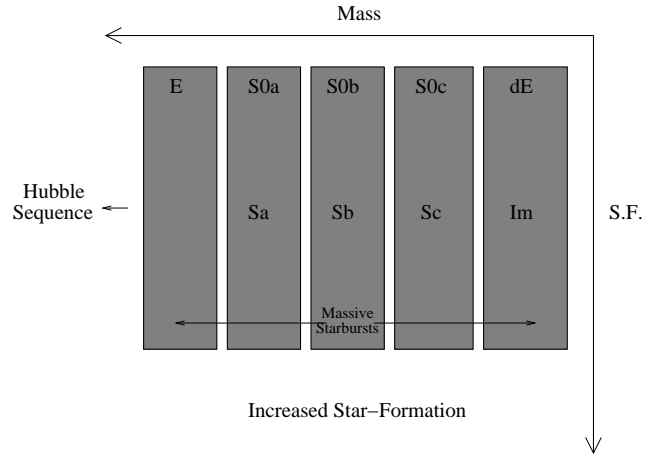}
 \caption{The system in Figure~11 projected onto the star formation/mass
plane.  The Hubble sequence is labelled, and includes galaxies with a range
of masses, modest star formation with no distinction made for 
interactions.  Ellipticals, S0s, dEs and dwarf spheroidals are located at
the bottom of the star formation sequence.  Massive star bursts are 
those galaxies located at the high star formation area of this projection.}
 \label{sample-figure}
\end{figure}

\begin{figure}
\includegraphics[width=84mm]{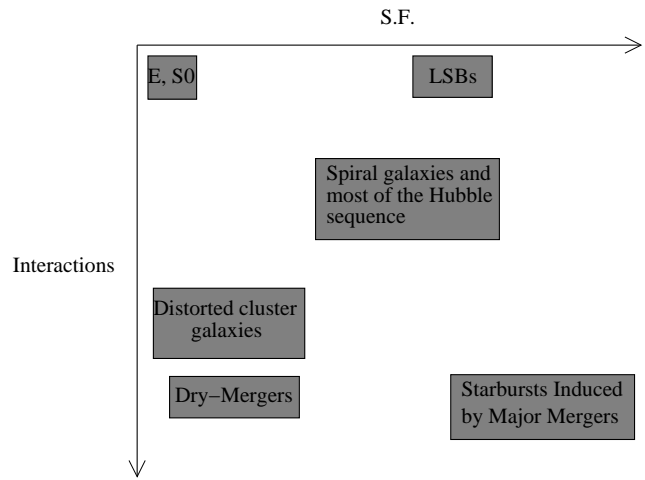}
 \caption{Projection of the star formation/interaction plane of Figure~11.
Several classes of galaxies separate along this plane perpendicular to
the mass sequence.  Low surface brightness galaxies (LSBs) are those with 
modest star formation and very little recent interaction history.  
Ellipticals and S0s are objects with relatively little star formation
or interactions, and no recent interactions/mergers.  
Several other galaxy types, such as distorted cluster galaxies undergoing
little star formation due to gas removal by the cluster, dry-mergers,
and starbursts induced by major mergers, are also labelled.}
 \label{sample-figure}
\end{figure}

This system is not completely new and various aspects of it have been
published in different forms. For example, van den Bergh (1976) proposed a 
revision of
the Hubble system where S0 galaxies occupy a parallel sequence to the 
spirals, instead of lodged between the ellipticals and Sa galaxies as in
the original Hubble sequence.  van den Bergh's (1976) classification
also contains an anaemic galaxy sequence for objects with little gas.  This
system also classifies galaxies in terms of their bulge to disk ratios,
making it very similar in some ways to the star formation/mass plane
presented here (see Figure~12).

Low surface brightness galaxies, as discussed earlier, have little
evidence for companions or signs of interactions with other galaxies.
They also have a very gradual star formation history that is still
ongoing.  As such they occupy the same mass sequence as the spirals,
but are located on the low interaction side of the interaction sequence.

The dichotomy between ellipticals that appear boxy and disky can also
be represented by this system (Kormendy \& Bender 1996), 
since these galaxies
naturally fall along different regions of interaction and mass space.  
Disky ellipticals, with rotation, would have a lower mass, while the 
degree of boxiness correlates with the interaction history.

Other galaxies can also be classified in this system.  Starbursts
induced by major mergers,
are placed at the high end of the interaction and
star formation sequence.  Galaxies undergoing violent relaxation, such as 
Arp 220, containing spheroidal components and step light profiles 
(Wright et al. 1990), would be classified 
towards the high-mass end of the mass sequence. 

This system is constructed based on the results presented in the earlier
part of this paper.  Measuring the star formation, merging, and
masses (either stellar or dynamical) properties of a galaxy are not
trivial, but are likely to become easier in the future.    We are
therefore not setting out in this section a stringent method for classifying 
galaxies by these properties.  One could do these classifications directly by
measuring the star formation rate through emission lines, masses through
dynamics or spectral energy distributions, and mergers/interactions
through an asymmetry index or a dynamical indicator.  We discuss in \S 6.3
a possible way to use the apparent morphologies of galaxies in optical
light to carry out these classifications.  Future work will reveal
better ways of carrying out these measurements, which can then be applied
for classification according to Figure~11.

\subsubsection{Other Systems as Projections}

Other morphological systems, including physical morphological ones, can
be viewed as projections of this system.  The Hubble sequence, to first
order, is a one
dimension projection with the star formation and interactions
collapsed onto the mass sequence.   As discussed before, the Hubble
sequence does not take into account these features, thus there exists a
wide range of properties at each Hubble type (Roberts \& Haynes
1994; Kennicutt 1998).   

The colour-asymmetry diagram is a projection, in part, of the star formation,
interaction sequence (Figure~13).  This is not an exact projection since the 
asymmetry parameter measures interactions, as well as star formation 
(Conselice et al. 2000a).  To first order the $(B-V)_{0}$ colour  
measures the star formation history of a galaxy.  Galaxies with purely
old stars have very red colours, while those with recent star formation are
bluer. 
The concentration-asymmetry diagram used by Bershady et al. (2000) and 
others, is furthermore a projection of the mass-interaction sequence.
More recently, Driver et al. (2006) argue that light concentration and
colour are fundamental properties that separate galaxies into early
and late types, which is again a two-dimensional projection of this
new system. Likewise, Blanton et al. (2005) suggest that colour and
and luminosity are the most important galaxy properties, which
is also shown in our analysis.

\subsubsection{Idealised Classifications}

When Hubble invented his classification, he did so with 
specific examples of galaxies, however Hubble types are now thought
to characterise ideal types.  There are few galaxies that fit this ideal, 
nor is any one galaxy exactly similar 
to any other.  This is due to the chaotic interaction and star 
formation history of galaxies that are, at present, not easily retrievable. 

The classification system presented here replaces ideal galaxies with
parameters that are in some ways idealised measurements.  It is not
trivial to determine the star formation history of a galaxy, or even
accurately measure its current star formation rate.  Star formation
does have a morphological appearance in most galaxies, the exception
being systems heavily enshrouded by dust.  The
system presented here can still account for those galaxies, as it is very
general. The method for classifying the star formation properties of a galaxy
does not have to be based on morphology, but can also be done 
through other properties, such as fluxes at various wavelengths.

The same is true for measuring the merger history of galaxies. While
many galaxies have some clues left over from merger events, these
clues can be quite diverse, requiring various methods to retrieve
them.   Structural methods, such as the asymmetry CAS parameter
(Conselice et al. 2000a,b), are useful for determining the 
presence of, and quantifying, current major mergers. 
Determining the history of past interactions is more difficult.  One way is
to investigate the outer parts of galaxies where the dynamical time
scales are long, and where apparently normal galaxies show evidence for
past merger activity (Schweizer \& Seitzer 1988).  Other methods include 
searching for distorted HI halos surrounding galaxies, galaxy core
depletion from merging black holes (Graham 2004), and  finding 
distorted rotation curves (Swaters et al. 1999) or line profiles 
(Haynes et al. 1998;
Conselice et al. 2000b).  These features are potentially all remnants of
accretion, merging or interactions with other galaxies, and can in
principle be used to quantify the interaction history of galaxies.

\subsubsection{Unidentified Galaxies?}

The system presented here can classify all known galaxies, but it
also has classification `corners' where no nearby galaxies are known to 
exist, or
have not been identified.  One type are massive, low interaction high star 
formation systems.  There are plenty of galaxies that have these
features at high redshift. These are likely ellipticals or disks in 
formation that will evolve into spheroids.  For example, systems with 
high mass, high star formation, and little merging could originate from gas 
accreted onto  elliptical-like systems.
Other possible galaxies not yet identified are  those with various mass
ranges, but low star formation and high interaction strength.  
Perhaps the population of distorted, but non-star forming
galaxies, or compact dwarfs found in clusters (Conselice \& Gallagher 1999;
Drinkwater et al. 2003) are examples of 
these, where the gas has been removed by the intracluster medium.
Another example are the 'dry' mergers between spheroids with little
gas (e.g., Hernandez-Toledo et al. 2006).
Other possible galaxy populations can be hypothesised by studying 
Figure~11.

\subsection{An Optical View of Galaxy Classification}

We have argued that the three parameters: mass, star formation and interactions
can be used together to classify and trace the evolution of galaxies. 
These parameters have a direct correlation with the semi-irreversible
processes of mass assembly or removal, and star formation where
gas mass is converted into stellar mass by some conversion efficiency $\alpha$.
Therefore, in principle these features can be measured through stellar
light distributions, which has been the primary method for
classifying galaxies for nearly a century.  We show in this section that this is possible.  The following description follows closely the more detailed
calculation given in appendix B of Conselice (2003).

We can write the initial baryonic mass of a galaxy after its
initial formation, (M${\rm _{0,B}}$) as,

\begin{equation}
{\rm M_{0,B}} = {\rm M_{0,S}} + {\rm M_{0,G}},
\end{equation}

\noindent where M$_{\rm 0,S}$ and M$_{\rm 0,G}$ are the initial stellar
and gas masses.  Over time, mass will accrete or be removed from a galaxy due 
to interactions and mergers.  At a time $T$, the amount
of mass added to or removed from the system, $M'(T)$ is given by,

\begin{equation}
M'(T) = \int^{T}_{0} \dot{M}(t) {\rm dt} = \int^{T}_{0} \dot{M}_{\rm S}(t) {\rm dt} + \int^{T}_{0} \dot{M}_{\rm G}(t) {\rm dt},
\end{equation}

\noindent with $\dot{M}$ the mass accretion, or removal, rate as a function of
time, and $T$ is the age of the galaxy when observed.  We can further divide 
the mass flux 
into stellar $\dot{M_{\rm S}}$ and gas $\dot{M_{\rm G}}$ components.
Some gas will be converted into stars at each
time, t, by a conversion rate, $\alpha(t)$, such that the mass converted
into stars, $\tilde{M_{\rm S}}$, is given by

\begin{equation}
\tilde{M_{\rm S}}(T) = \int^{T}_{0} \alpha(t)M_{\rm G}(t) {\rm dt},
\end{equation}

\noindent where $\alpha(t)$ and $M_{\rm G}$ are the gas conversion rate,
and gas mass as a function of time.  We can then write the total stellar and 
gas masses as a function of time by:

\begin{equation}
M_{\rm S}(t) = {\rm M_{0,S}} + \int^{T}_{0} \left(\dot{M}_{\rm S}(t) + \alpha(t)M_{\rm G}(t)\right) {\rm dt},
\end{equation}

\begin{equation}
M_{\rm G}(t) = {\rm M_{0,G}} + \int^{T}_{0} \left(\dot{M}_{\rm G}(t) - \alpha(t)M_{\rm G}(t)\right) {\rm dt}.
\end{equation}

\noindent The three axes of our system (Figure~11) can be represented
by three terms in these equations. An interaction
parameter can be represented by $M'_{\rm S}(t)$, and a star formation 
parameter by $\tilde{M_{\rm S}}$. We can also
create dimensionless parameters that in principal  correlate with measurable
stellar light features.  We can write the integrated interaction, $\bar{I(t)}$
and star formation $\bar{F(t)}$ strengths weighted by the 
final mass, and time as:

\begin{equation}
\bar{I(t)} = \int^{T}_{0} \frac{t}{T} \frac{\left|\dot{M}_{\rm S}(t)\right|}{M_{\rm S}(T)} {\rm dt}
\end{equation}

\begin{equation}
\bar{F(t)} = \int^{T}_{0} \frac{t}{T} \frac{\alpha(t) M_{\rm G}(t)}{M_{\rm S}(T)} {\rm dt}.
\end{equation}

\noindent These two ideal morphological parameters are 
quantities that should correlate with the appearance of a galaxy, and are
further elaborated in appendix B of Conselice (2003).  They 
are weighted by time since older star formation and interactions will
have less of an effect on the appearance of a galaxy.  Likewise, more massive
galaxies will have appearances that, in general, will be less affected by
relatively lower amounts of gas converted into stars, and stars lost or 
gained from galaxy interactions.

What this section shows is that the main galaxy formation processes - star
formation and interactions/mergers can be traced through optical light. 
Typically, a galaxy which has undergone recent star formation will contain
clumpy light, occupied by clusters of young stars.  Likewise, a galaxy
undergoing a merger or interaction will appear lopsided and/or distorted
in some manner. The mass of a galaxy often correlates with the
light concentration, which can be estimated by eye.  Therefore, by
estimating the degree of: disturbances, clumpy structures, and how 
concentrated a galaxy
is (i.e., bulge vs. disk structures) it is possible to classify galaxies
into this new system using optical light.A detailed exploration of this 
idea in terms of eye-ball classifications is however beyond the scope of 
this current paper.  Quantitatively, this approach has been
successfully used to classify both nearby (Conselice et al. 2002; Conselice 
2003) and high redshift galaxy populations (Conselice et al. 2003b; 
Conselice et al. 2004; Grogin et al. 2005; Papovich et al. 2005; Lehmer et al. 2005; Conselice et al. 2005c; Lotz et al. 2006) within the CAS system.  

\section{Summary and Conclusions}

This papers attempts to answer a few very basic questions about galaxies
and their properties.  The first is how galaxies
of different morphological types are related to each other, including
the relationship between various major evolutionary processes.  
Another question we
address is whether or not the structural appearance of a galaxy contains
enough information to offer this classification.   To answer the
first question, which constitutes the bulk of the paper, we carried out a 
series of statistical analyses,
including principal component analyses, of fourteen properties in
22,121 galaxies.  The major results of our investigations are:

\noindent I.  Spearman correlation analysis show that most measurable
properties of galaxies are correlated with each other, and that Hubble
types correlate strongest with stellar mass and $(U-B)_{0}$ and $(B-V)_{0}$
colour.  Scale features (M$_{*}$, M$_{\rm B}$, radius, HI gas content,
far-infrared luminosity) correlate strongly with each other independent
of morphological, colour, or luminosity selection.  Furthermore, the
strong correlation between M$_{*}$ and colour is independent of Hubble
type.

\noindent II.  We carry out several principal component analyses (PCA)
to determine how many PCA eigenvectors are needed to describe the major
properties of all nearby galaxies.  Our conclusion from this is that
the major eigenvector is dominated by the scale, or mass, of a galaxy, while
the next two major eigenvectors are dominated by the star formation, and
galaxy interactions/mergers, respectively.

\noindent III. Based on points I and II, we conclude that the three major 
processes that should be used to classify a galaxy are: the scale or mass, 
interactions and mergers with other galaxies, and recent star-formation.

\noindent IV. We argue that these three features (mass, star formation
and interactions) are revealed through the stellar mass distributions within
galaxies. Thus the optical structure of a galaxy can be used, in
principle, to accurately classify galaxies into a physically meaningful
system.  One way this can be done is through the CAS parameters described in 
Conselice (2003).

\noindent V. Based on our analysis we present a morphological
break down of $z \sim 0$ galaxies, including the fraction and
number densities of different
Hubble types (ellipticals, spirals, irregulars) at different luminosities.
We find that the bulk of nearby bright galaxies with M$_{\rm B} < -20$
are spirals (69\%), while 19\% are elliptical, 10\% S0 and
2\% irregular. We also investigate which types of galaxies are barred, 
and the number densities
of barred galaxies in the nearby universe. These are intended to be useful
for higher redshift comparisons, which are just now becoming possible.

In summary, this paper presents a new basic methodology for understanding and
classifying 
galaxies. It is a physical system that uses the quantifiable parameters of 
mass, star formation, and interactions, and is not based on galaxy
morphology.  The system can account for all known galaxy types, and is a 
naturally outgrowth of detailed studies of galaxies over the past century.   
Future work on this classification scheme will include understanding
the origin of the mass
sequence, and why it correlates at $z \sim 0$ with star formation and
bulge to disk ratios. Furthermore, our new classification system is by no means
finalised. In the future, we will likely be able to measure galaxy
masses, star formation rates, and ongoing interactions with a high
accuracy.  Whether these measurements are done morphologically or not
is irrelevant, as these three properties measured in any way can be
used to classify galaxies.

\section*{Acknowledgements}

This work, in one way or another, has been in progress since 1995, although
the bulk of it was completed and written in 2001-2002 and 2006.  I thank
the University of Wisconsin-Madison, the Space Telescope Science
Institute, Caltech and the University of Nottingham for their
support while this work was completed.  An early form of section 6 
was originally included 
in the conclusions to my PhD thesis.  I thank Jay
Gallagher and Matt Bershady for their collaboration which led to many
of the ideas discussed in this paper, and Alister Graham, 
Michael Merrifield and
Sidney van den Bergh for reading and commenting on various drafts. I
also thank the referee for a very constructive report with many 
useful suggestions.   Finally, I thank Sylvia and
Scott Mencner for the inspiration to produce Figure~11.  Financial
support for this work was provided by the National Science Foundation 
through
a Astronomy and Astrophysics postdoctoral Fellowship, NASA, and the 
UK Particle Physics and Astronomy Research Council through
grants to the University of Nottingham.  This research has made use of 
the Lyon extragalactic Database (LEDA) and the NASA/IPAC Extragalactic 
Database (NED) which is operated by the Jet 
Propulsion Laboratory, California Institute of Technology, under contract 
with the National Aeronautics and Space Administration.

\label{lastpage}

\end{document}